\documentclass[11pt,a4paper]{article}

\usepackage{bm,amsfonts}
\usepackage{graphicx}
\usepackage{hyperref}
\usepackage{youngtab}

\newcommand{\Z}{\mathbb{Z}}
\newcommand{\C}{\mathbb{C}}
\newcommand{\R}{\mathbb{R}}
\newcommand{\N}{\mathcal{N}}

\newcommand{\PV}{\mathrm{P}}
\newcommand{\U}{\mathrm{U}}
\newcommand{\SU}{\mathrm{SU}}

\makeatletter

\@addtoreset{equation}{section}
\makeatother

\date{\today}

\begin{document}

\begin{titlepage}

\renewcommand{\thefootnote}{\fnsymbol{footnote}}

\begin{flushright}
RIKEN-MP-21
\\
\end{flushright}

\vskip5em

\begin{center}
 {\Large {\bf 
 Matrix model from $\N=2$ orbifold partition function
 }}

 \vskip3em

 {\sc Taro Kimura}\footnote{E-mail address: 
 \href{mailto:kimura@dice.c.u-tokyo.ac.jp}
 {\tt kimura@dice.c.u-tokyo.ac.jp}}

 \vskip2em

{\it Department of Basic Science, University of Tokyo, 
 Tokyo 153-8902, Japan\\ \vskip.2em
 and\\ \vskip.2em
 Mathematical Physics Lab., RIKEN Nishina Center, Saitama 351-0198,
 Japan 
}

 \vskip3em

% \today
\end{center}

 \vskip2em

\begin{abstract}
 The orbifold generalization of the partition function, which would
 describe the gauge theory on the ALE space, is investigated from the
 combinatorial perspective.
 It is shown that the root of unity limit $q \to \exp(2\pi i/k)$ of the
 $q$-deformed partition function plays a crucial role in the orbifold
 projection while the limit $q \to 1$ applies to $\R^4$.
 Then starting from the combinatorial representation of the partition
 function, a new type of multi-matrix model is derived by considering its
 asymptotic behavior.
 It is also shown that Seiberg-Witten curve for the corresponding gauge
 theory arises from the spectral curve of this multi-matrix model.

\if0
 Keywords:
 instanton counting for the ALE spaces, Moduli spaces of instantons on
 the ALE spaces, affine Lie algebra and quiver varieties...
 \fi
\end{abstract}

\end{titlepage}

\tableofcontents

\setcounter{footnote}{0}

%%%%%%%%%%   Introduction   %%%%%%%%%%

\section{Introduction}
\label{sec:Intro}

The recent progress on the four dimensional $\N=2$
supersymmetric gauge theory reveals a remarkable relation to the two
dimensional conformal field theory \cite{Alday:2009aq}.
This relation provides the
explicit interpretation for the partition function of the four
dimensional gauge theory \cite{Nekrasov:2002qd,Nekrasov:2003rj} as the
conformal block of the two dimensional Liouville field theory, and is
naturally regarded as a consequence of the M-brane compactifications
\cite{Witten:1997sc,Gaiotto:2009we}, which reproduces the results of the four
dimensional gauge theory \cite{Seiberg:1994rs,Seiberg:1994aj}.
It is originally considered in \cite{Alday:2009aq} for $\SU(2)$ theory, and
extended to the higher rank gauge theory
\cite{Wyllard:2009hg,Mironov:2009by}, the non-conformal theory
\cite{Gaiotto:2009ma,Marshakov:2009gn,Taki:2009zd} and the cases with the
surface and loop operators \cite{Alday:2009fs}, etc.

According to this connection, established results on the two dimensional
side can be reconsidered from the viewpoint of the four dimensional
theory, and vice versa.
One of the useful applications is the matrix model description of the
supersymmetric gauge theory
\cite{Dijkgraaf:2009pc,Itoyama:2009sc,Eguchi:2009gf,Schiappa:2009cc,Itoyama:2010ki,Eguchi:2010rf,Itoyama:2010na,Itoyama:2011mr}.
This is based on the fact that the conformal block on the sphere can be also
regarded as the matrix integral, which is called the Dotsenko-Fateev
integral representation \cite{Dotsenko:1984nm,Dotsenko:1984ad}.
In this direction some extensions of the matrix model description are
performed by starting with the Liouville correlators on the higher genus
Riemann surfaces \cite{Maruyoshi:2010pw,Bonelli:2010gk}.
Furthermore another type of the matrix model is also investigated so
far \cite{Klemm:2008yu,Sulkowski:2009br,Sulkowski:2009ne} \cite{Tai:2007vc}.
This matrix model is directly derived from the combinatorial representation of
the partition function by considering its asymptotic behavior while the
Dotsenko-Fateev type matrix model is obtained from the Liouville
correlator, so that the conformal symmetry is manifest.
This treatment is quite analogous to the matrix integral representation
of the combinatorial object.\footnote{For example, the longest increasing
subsequences in random permutations, the non-equilibrium stochastic
model, so-called TASEP \cite{springerlink:10.1007/s002200050027}, and so
on (see also \cite{1742-5468-2007-07-P07007}).
Their remarkable connection to the Tracy-Widom
distribution \cite{springerlink:10.1007/BF02100489} can be understood
from the viewpoint of the random matrix theory through the
Robinson-Schensted-Knuth (RSK) correspondence (see e.g. \cite{Stanley200102}).}
Although they are apparently different from the Dotsenko-Fateev type,
both types of the matrix model correctly reproduce the results of the
four dimensional gauge theory, e.g. Seiberg-Witten curve.

The purpose of this paper is to extend the remarkable connection between
the two and four dimensional theory to the orbifold theory.
The four dimensional orbifold manifold is given
by $\C^2/\Gamma$ where $\Gamma$ is a finite subgroup of
$\SU(2)$, e.g. $\Gamma=\Z_k$, and its minimal resolution of the
singularity gives the ALE space \cite{Kronheimer:1989zs}.
The gauge theory on the ALE space is well investigated in
\cite{springerlink:10.1007/BF01233429,springerlink:10.1007/BF01444534}
with respect to the instanton moduli space.
Their results show the moduli spaces of the instanton on the ALE spaces are
deeply connected to the affine Lie algebras, and the quiver varieties.
This interesting fact provides a guideline to search for a counterpart
of the AGT relation for the orbifold theory.
Furthermore this theory has been reconsidered in terms of D-branes
\cite{Dijkgraaf:2007sw} in which some aspects of 2d/4d connection are
partially studied, and the theory on the Taub-NUT space has been
investigated in \cite{Witten:2009xu} in detail.

To obtain the solutions of the four dimensional gauge theory on the ALE
space, we have to deal with the partition function as the case of
$\R^4$ by implementing the ADHM construction and the
localization method for the ALE space. 
By performing such a procedure, the orbifold generalization of the
combinatorial partition function, which would describe the gauge theory
on the ALE space, has been investigated \cite{Fucito:2004ry} (see also
\cite{Dijkgraaf:2007fe}).
They are defined as the invariant sector of the Young diagram under the
orbifold action $\Gamma$, and the process to extract only the
$\Gamma$-invariant sector is called the orbifold projection in
\cite{Fucito:2004ry}.
This projection seems quite reasonable, but it is difficult to perform
this in a systematic manner.
In this paper we will show the orbifold projection is naturally
performed when one appropriately parametrizes the partition function.
This parametrization is just interpreted as the root of unity limit $q
\to \exp\left(2\pi i/k\right)$ of the $q$-deformed partition function
while we have to take $q \to 1$ for the usual $\R^4$ theory.\footnote{
Because the combinatorial partition function includes infinite products
of the $q$-parameter, we have to take care of its convergence radius.
Thus, to obtain the root of unity limit $q\to\exp(2\pi i/k)$, we first
parametrize it as $q \to \omega q$, and take the limit of $q\to 1$.
}
The similar approach is found in \cite{Uglov:1997ia} in the context of
the spin Calogero-Sutherland model, where the combinatorial method plays
an important role in characterization of the wavefunction, derivation of
the dynamical correlation functions and so on (see, for example,
\cite{KuramotoKato200908}). 

In this paper we also propose a new matrix model description for the
orbifold theory.
Starting with the combinatorial representation and applying the method
developed in \cite{Klemm:2008yu,Sulkowski:2009br,Sulkowski:2009ne} we
derive a new kind of matrix models, and discuss the corresponding
gauge theory consequences.
As a result of orbifolding, we get $k$-matrix models for the ALE space.
This model is quite similar to the Chern-Simons matrix
model \cite{Marino:2002fk} for the lens space $S^3/\Z_k$
\cite{Aganagic:2002wv,Halmagyi:2003ze,Halmagyi:2003mm} (see also
\cite{Marino:2011nm}), which is
recently applied to the ABJM theory \cite{Aharony:2008ug,Aharony:2008gk}
\cite{Kapustin:2010xq,Drukker:2010nc}.
This is so reasonable because the ALE space
goes to the lens space $S^3/\Z_k$ at infinity.
We then discuss the large $N$ limit of the matrix model, and show
Seiberg-Witten curve is arising from the spectral curve of the matrix model.

Along the recent interest in the 2d/4d relation, it is natural to search
a two dimensional counterpart of the orbifold partition function.
While the original relation relies on the interpretation of the $\SU(n)$
partition function as the conformal block of the conformal field theory
described by $W_n$ algebra, its generalization is also proposed
\cite{Alday:2010vg,Kozcaz:2010yp} \cite{Tachikawa:2011dz}
\cite{Kanno:2011fw}: the generalized $W$ algebra
corresponds to more complicated gauge theories in four dimensions.
In such a theory an embedding $\rho: \SU(2) \to \SU(n)$
plays an important role in characterizing the generalized $W$ algebra
\cite{deBoer:1993iz}.
On the other hand, in the orbifold theory, an embedding $\rho: \Gamma
\subset \SU(2) \to \SU(n)$ is also found to characterize the
decomposition into the irreducible representations of $\Gamma$.
We will discuss this similarity from the viewpoint of string theory and
propose a new kind of connection between 2d/4d theories.

The organization of this paper is as follows.
In section \ref{sec:N=2} we provide the combinatorial representation of
the four dimensional gauge theory, which is originally proposed in
\cite{Nekrasov:2002qd,Nekrasov:2003rj}, as preliminaries for some
extensions discussed in this paper.
In section \ref{sec:orbifold} we then study the orbifold generalization
of the partition function with focusing on its combinatorial aspect.
We will see the root of unity limit of the $q$-deformed partition
function implements the orbifold projection.
Section \ref{sec:matrix} is devoted to derivation of the matrix model
from the orbifold partition function by considering its asymptotic behavior.
In section \ref{sec:spectral} we study the multi-matrix model in detail
by taking the large $N$ limit. We then extract
Seiberg-Witten curve of the corresponding gauge theory via the spectral
curve of the matrix model.
In section \ref{sec:related} we try to search a 2d/4d relation for the
orbifold theory.
The two dimensional theory is discussed from the viewpoint of string
theory, and a relation to the generalized $W$ algebra is also proposed.
We also discuss topics related to the $q$-deformation of the
partition function.
We finally summarize our results in section \ref{sec:summary}.

%%%%%%%%%  Note added  %%%%%%%%%

\subsubsection*{Note added}

After submitting this article, some related
papers, considering extensions of the AGT relation to the ALE spaces,
appear in the preprint server
\cite{Belavin:2011pp,Nishioka:2011jk,Bonelli:2011jx,Belavin:2011tb,Bonelli:2011kv}.

\section{$\N=2$ partition functions}\label{sec:N=2}

Let us start with the partition functions for $\N=2$
theories, originally studied by \cite{Nekrasov:2002qd,Nekrasov:2003rj},
as preliminaries of discussions along this paper: its deformation is proposed
in section \ref{sec:orbifold} and we will also see in section
\ref{sec:matrix} the matrix model can be obtained from the combinatorial
representation by considering its asymptotic behavior.

First we introduce the instanton part of  the four dimensional partition
function for $\N=2$ $\SU(n)$ theory with $N_f$
(anti)fundamental matters,
\begin{eqnarray}
 Z^{\rm 4d} & = & \sum_{\vec{\lambda}} \Lambda^{(2n-N_f)|\vec \lambda|}
  Z_{\rm vec}^{\rm 4d} Z_{\rm (anti)fund}^{\rm 4d}, \\
 Z_{\rm vec}^{\rm 4d} & = & \prod_{(l,i)\not=(m,j)}
  \frac{\Gamma(\lambda_i^{(l)}-\lambda_j^{(m)}+\beta(j-i)+b_{lm}+\beta)}
       {\Gamma(\lambda_i^{(l)}-\lambda_j^{(m)}+\beta(j-i)+b_{lm})}
  \frac{\Gamma(\beta(j-i)+b_{lk})}{\Gamma(\beta(j-i)+b_{lk}+\beta)} ,
  \nonumber \\
 && \\
 Z_{\rm fund}^{\rm 4d} & = & 
  \prod_{l=1}^n \prod_{f=1}^{N_f} \prod_{i=1}^\infty
  \frac{\Gamma(\lambda_i^{(l)}+b_l+M_f+\beta i+1)}
       {\Gamma(b_l+M_f+\beta i+1)}, \\
 Z_{\rm antifund}^{\rm 4d} & = & 
  \prod_{l=1}^n \prod_{f=1}^{N_f} \prod_{i=1}^\infty
  \frac{\Gamma(\lambda_i^{(l)}+b_l-M_f+\beta (i+1))}
       {\Gamma(b_l-M_f+\beta (i+1))}.
\end{eqnarray}
These combinatorial expressions are written in terms of $n$-tuple
partitions, $\vec \lambda=(\lambda^{(1)}, \cdots, \lambda^{(n)})$, and
their parameters are related to those of the gauge theory as
\begin{equation}
 \beta = - \frac{\epsilon_1}{\epsilon_2}, \qquad
 b_l = \frac{a_l}{\epsilon_2}, \qquad
 a_{lm} = a_l - a_m, \qquad
 b_{lm} = b_l - b_m, \qquad
 M_f = \frac{m_f}{\epsilon_2}
\end{equation}
where $\epsilon_1 > 0 > \epsilon_2$ are $\Omega$-background parameters,
$a_l$ is the Coulomb moduli parametrizing the $\U(1)^{n-1}$ vacua, and $m_f$
denotes the mass of the fundamental matter.
We can also introduce bifundamental matter fields, but we do not focus
on them in this paper.
For $\N=2^*$ theory, which includes the adjoint matter, we have
to consider another contribution,
\begin{equation}
  Z_{\rm adj}^{\rm 4d} = \prod_{(l,i)\not=(m,j)}
  \frac{\Gamma(\lambda_i^{(l)}-\lambda_j^{(m)}
        +\beta(j-i)+b_{lm}+\mathfrak{M})}
       {\Gamma(\lambda_i^{(l)}-\lambda_j^{(m)}
        +\beta(j-i)+b_{lm}+\mathfrak{M}+\beta)}
  \frac{\Gamma(\beta(j-i)+b_{lk}+\mathfrak{M}+\beta)}
       {\Gamma(\beta(j-i)+b_{lk}+\mathfrak{M})}
\end{equation}
with the mass of the adjoint matter
\begin{equation}
 \mathfrak{M} = \frac{\mathfrak{m}}{\epsilon_2}. 
\end{equation}
Note one can see a simple relation
\begin{equation}
 Z_{\rm vec} = \frac{1}{Z_{\rm adj}(\mathfrak{m}=0)}.
\end{equation}
This $\N=2^*$ theory is given by the simple deformation of
$\N=4$ theory by applying the mass to the adjoint matter.
We will comment on the massless limit of this $\N=2^*$ theory later.

The five dimensional extensions of these partition
functions are also proposed,
which can be regarded as the $q$-deformed analogue of the original four
dimensional one,
\begin{equation}
 Z^{\rm 5d} = \sum_{\vec{\lambda}} \Lambda^{(2n-N_f)|\vec \lambda|}
  Z_{\rm vec}^{\rm 5d} Z_{\rm (anti)fund}^{\rm 5d},
\end{equation}
\begin{eqnarray}
 Z^{\rm 5d}_{\rm vec} & = & 
  \prod_{(l,i)\not=(m,j)}
  \frac{(Q_{lm}q^{\lambda_i^{(l)}-\lambda_j^{(m)}}t^{j-i};q)_\infty}
       {(Q_{lm}q^{\lambda_i^{(l)}-\lambda_j^{(m)}}t^{j-i+1};q)_\infty}
  \frac{(Q_{lm}t^{j-i+1};q)_\infty}
       {(Q_{lm}t^{j-i};q)_\infty},
 \label{part_vec01} \\
 Z_{\rm fund}^{\rm 5d} & = & 
  \prod_{l=1}^n \prod_{f=1}^{N_f} \prod_{i=1}^\infty
  \frac{(Q_l Q_{m_f}qt^{-i};q)_\infty}
       {(Q_l Q_{m_f}q^{\lambda_i^{(l)}+1}t^{-i};q)_\infty} , 
 \label{part_fund01} \\
 Z_{\rm antifund}^{\rm 5d} & = & 
  \prod_{l=1}^n \prod_{f=1}^{N_f} \prod_{i=1}^\infty
  \frac{(Q_l Q_{m_f}^{-1} t^{-i+1};q)_\infty}
       {(Q_l Q_{m_f}^{-1} q^{\lambda_i^{(l)}}t^{-i+1};q)_\infty},
 \label{part_fund02} \\
 Z_{\rm adj}^{\rm 5d} & = & \prod_{(l,i)\not=(m,j)}
  \frac{(Q_{\mathfrak{m}}Q_{lm}
         q^{\lambda_i^{(l)}-\lambda_j^{(m)}}t^{j-i+1};q)_\infty}
       {(Q_{\mathfrak{m}}Q_{lm}
         q^{\lambda_i^{(l)}-\lambda_j^{(m)}}t^{j-i};q)_\infty}
  \frac{(Q_{\mathfrak{m}}Q_{lm}t^{j-i};q)_\infty}
       {(Q_{\mathfrak{m}}Q_{lm}t^{j-i+1};q)_\infty} .
 \label{part_adj01}
\end{eqnarray}
Here we define
\begin{equation}
 (x;q)_\infty = \prod_{p=0}^\infty (1-xq^p),
\end{equation}
and deformed parameters are given by
\begin{equation}
 q = e^{\epsilon_2}, \qquad
 t = e^{-\epsilon_1} = q^\beta, \qquad
 Q_l = e^{a_l} = q^{b_l}, \qquad
 Q_{lm} = e^{a_l - a_m} = q^{b_{lm}},
\end{equation}
\begin{equation}
 Q_{m_f} = e^{m_f} = q^{M_f}, \qquad
 Q_{\mathfrak{m}} = e^{\mathfrak{m}} = q^{\mathfrak{M}}.
\end{equation}
The radius $R$ of the compactified dimension $S^1$ is implicitly
included in these parameters by rescaling the parameters,
e.g. $q=e^{R\epsilon_2}$.
Thus one can check the four dimensional result is reproduced by the five
dimensional partition function by taking the limit $R \to 0$, or
equivalently $q \to 1$.

To obtain the matrix models, it is useful to consider another
representation for these partition functions with cut-off parameters
$N^{(l)}$ for the partitions
\cite{Klemm:2008yu,Sulkowski:2009br,Sulkowski:2009ne}.
By decomposing the partition functions (\ref{part_vec01}),
(\ref{part_adj01}) as
\begin{eqnarray}
  Z_{\rm vec} & = & \prod_{l,m}^n Z_{\rm vec}^{(l,m)},
   \label{part_vec02} \\
 Z_{\rm adj} & = & \prod_{l,m}^n Z_{\rm adj}^{(l,m)},
\end{eqnarray}
each part can be rewritten as
\begin{eqnarray}
 Z_{\rm vec}^{(l,l)} & = & \prod_{i\not=j}^\infty
  \frac{(q^{\lambda_i^{(l)}-\lambda_j^{(l)}}t^{j-i};q)_\infty}
       {(q^{\lambda_i^{(l)}-\lambda_j^{(l)}}t^{j-i+1};q)_\infty}
  \frac{(t^{j-i+1};q)_\infty}
       {(t^{j-i};q)_\infty}
 \nonumber \\
 & = & \frac{1}{A_{\mathfrak{m}=0}^{(l,l)}} 
  \prod_{i\not=j}^{N^{(l)}}
  \frac{(q^{\lambda_i^{(l)}-\lambda_j^{(l)}}t^{j-i};q)_\infty}
       {(q^{\lambda_i^{(l)}-\lambda_j^{(l)}}t^{j-i+1};q)_\infty}
  \prod_{i=1}^{N^{(l)}}
  \frac{(q^{\lambda_i^{(l)}}t^{N^{(l)}-i+1};q)_\infty}
       {(q^{-\lambda_i^{(l)}}t^{-N^{(l)}+i};q)_\infty} ,
 \\
 Z_{\rm vec}^{(l,m)} & = & \prod_{i,j}^\infty
  \frac{(Q_{lm}q^{\lambda_i^{(l)}-\lambda_j^{(m)}}t^{j-i};q)_\infty}
       {(Q_{lm}q^{\lambda_i^{(l)}-\lambda_j^{(m)}}t^{j-i+1};q)_\infty}
  \frac{(Q_{lm}t^{j-i+1};q)_\infty}
       {(Q_{lm}t^{j-i};q)_\infty}
 \nonumber \\
 & = & \frac{1}{A_{\mathfrak{m}=0}^{(l,m)}} 
  \prod_{i=1}^{N^{(l)}} \prod_{j=1}^{N^{(m)}}
  \frac{(Q_{lm}q^{\lambda_i^{(l)}-\lambda_j^{(m)}}t^{j-i};q)_\infty}
       {(Q_{lm}q^{\lambda_i^{(l)}-\lambda_j^{(m)}}t^{j-i+1};q)_\infty}  
  \frac{\prod_{i=1}^{N^{(l)}}
        (Q_{lm}q^{\lambda_i^{(l)}}t^{N^{(m)}-i+1};q)_\infty}
       {\prod_{i=1}^{N^{(m)}}
        (Q_{lm}q^{-\lambda_i^{(m)}}t^{-N^{(l)}+i};q)_\infty},
 \\
 Z_{\rm adj}^{(l,l)} & = & {A^{(l)}_{\mathfrak{m}}} \prod_{i\not=j}^{N^{(l)}}
  \frac{(Q_{\mathfrak{m}}q^{\lambda_i^{(l)}-\lambda_j^{(l)}}t^{j-i+1};q)_\infty}
       {(Q_{\mathfrak{m}}q^{\lambda_i^{(l)}-\lambda_j^{(l)}}t^{j-i};q)_\infty}
  \prod_{i=1}^{N^{(l)}}
  \frac{(Q_{\mathfrak{m}}q^{-\lambda_i^{(l)}}t^{-N^{(l)}+i};q)_\infty}
       {(Q_{\mathfrak{m}}q^{\lambda_i^{(l)}}t^{N^{(l)}-i+1};q)_\infty} ,
 \\
 Z_{\rm adj}^{(l,m)} & = & {A^{(l,m)}_{\mathfrak{m}}} \prod_{i=1}^{N^{(l)}} \prod_{j=1}^{N^{(m)}}
  \frac{(Q_{\mathfrak{m}}Q_{lm}q^{\lambda_i^{(l)}-\lambda_j^{(m)}}t^{j-i+1};q)_\infty}
       {(Q_{\mathfrak{m}}Q_{lm}q^{\lambda_i^{(l)}-\lambda_j^{(m)}}t^{j-i};q)_\infty}
  \frac{\prod_{i=1}^{N^{(m)}}
        (Q_{\mathfrak{m}}Q_{lm}q^{-\lambda_i^{(m)}}t^{-N^{(l)}+i};q)_\infty}
       {\prod_{i=1}^{N^{(l)}}
        (Q_{\mathfrak{m}}Q_{lm}q^{\lambda_i^{(l)}}t^{N^{(m)}-i+1};q)_\infty} .
 \nonumber \\	
\end{eqnarray}
We can check finite constants turn out to be
\begin{eqnarray}
  {A^{(l,m)}_{\mathfrak{m}}} & = & 
   \left\{
    {\renewcommand\arraystretch{1.5} 
    \begin{array}{cl}
     {\displaystyle 
      \left(
       \frac{(Q_{\mathfrak{m}}t;q)_\infty}{(Q_{\mathfrak{m}};q)_\infty}
      \right)^{N^{(l)}}} & (l = m) \\
     {\displaystyle 1} & (N^{(l)} = N^{(m)}, l\not=m) \\
     {\displaystyle \prod_{i=1}^{N^{(m)}-N^{(l)}} (Q_{\mathfrak{m}}Q_{lm}t^i;q)_\infty^{-1}} 
      & (N^{(l)} < N^{(m)}) \\
     {\displaystyle \prod_{i=1}^{N^{(l)}-N^{(m)}} (Q_{\mathfrak{m}}Q_{lm}t^{-i+1};q)_\infty }
     & (N^{(l)} > N^{(m)})
    \end{array}
    }
   \right. .
\end{eqnarray}
Therefore they can be represented as
\begin{eqnarray}
 Z_{\rm vec} & = & 
  \left(
   \prod_{l,m}^n \frac{1}{A^{(l,m)}_{\mathfrak{m}=0}}
  \right)
  \prod_{(l,i)\not=(m,j)}
  \frac{(Q_{lm}q^{\lambda_i^{(l)}-\lambda_j^{(m)}}t^{j-i};q)_\infty}
       {(Q_{lm}q^{\lambda_i^{(l)}-\lambda_j^{(m)}}t^{j-i+1};q)_\infty}
  \prod_{l,m}^n \prod_{i=1}^{N^{(l)}}
  \frac{(Q_{lm}q^{\lambda_i^{(l)}}t^{N^{(m)}-i+1};q)_\infty}
       {(Q_{ml}q^{-\lambda_i^{(l)}}t^{-N^{(m)}-i};q)_\infty},
 \\
 Z_{\rm adj} & = & 
  \left(
   \prod_{l,m}^n A^{(l,m)}_{\mathfrak{m}}
  \right)
  \prod_{(l,i)\not=(m,j)}
  \frac{(Q_{\mathfrak{m}}Q_{lm}
         q^{\lambda_i^{(l)}-\lambda_j^{(m)}}t^{j-i+1};q)_\infty}
       {(Q_{\mathfrak{m}}Q_{lm}
         q^{\lambda_i^{(l)}-\lambda_j^{(m)}}t^{j-i};q)_\infty}
  \prod_{l,m}^n \prod_{i=1}^{N^{(l)}}
  \frac{(Q_{\mathfrak{m}}Q_{ml}
         q^{-\lambda_i^{(l)}}t^{-N^{(m)}-i};q)_\infty}
       {(Q_{\mathfrak{m}}Q_{lm}
         q^{\lambda_i^{(l)}}t^{N^{(m)}-i+1};q)_\infty},
 \nonumber \\	
\end{eqnarray}
\begin{eqnarray}
 Z_{\rm fund} & = & 
  \prod_{l=1}^n \prod_{f=1}^{N_f} \prod_{i=1}^{N^{(l)}}
  \frac{(Q_l Q_{m_f} qt^{-i};q)_\infty}
       {(Q_l Q_{m_f} q^{\lambda_i^{(l)}+1}t^{-i};q)_\infty} , 
 \label{part_fund03} \\
 Z_{\rm antifund} & = & 
  \prod_{l=1}^n \prod_{f=1}^{N_f} \prod_{i=1}^{N^{(l)}}
  \frac{(Q_l Q_{m_f}^{-1} t^{-i+1};q)_\infty}
       {(Q_l Q_{m_f}^{-1} q^{\lambda_i^{(l)}}t^{-i+1};q)_\infty}.
 \label{part_fund04}
\end{eqnarray}
These expressions are convenient to give the matrix model description
\cite{Klemm:2008yu,Sulkowski:2009br,Sulkowski:2009ne}.
We will investigate deformed versions of the partition functions,
starting from these expressions, and derive the corresponding matrix
models in section \ref{sec:matrix}.

\section{Orbifold partition function}\label{sec:orbifold}

We then consider the orbifold generalization of the partition function
for $\N=2$ theory \cite{Fucito:2004ry}, which
describes the gauge theory on the ALE spaces
\cite{Kronheimer:1989zs,springerlink:10.1007/BF01233429,springerlink:10.1007/BF01444534}, obtained from the minimal resolution of orbifolds
$\C^2/\Gamma$ where $\Gamma$ is a finite subgroup of $\SU(2)$.
This resolution of the singularity is performed by replacing
two-spheres, where their intersecting numbers are related to a Cartan
matrix of the corresponding Lie algebras $\mathfrak{g}$.
This connection between subgroups $\Gamma$ of $\SU(2)$ and Lie algebras
$\mathfrak{g}$ is known as the McKay correspondence.
Especially the Abelian groups $\Gamma=\Z_k$ correspond to
$A_{k-1}$ Lie algebras, and we focus on them in this paper.

To study such an extended partition function, we will show it is useful
to deal with the $q$-deformed partition function which corresponds to
the five dimensional theory, and take the root of unity limit of the
deformation parameter.
We remark the procedure used in this paper to obtain the orbifold partition
function is closely related to the method proposed in
\cite{Uglov:1997ia} to study the
spin Calogero-Sutherland model (e.g. see \cite{KuramotoKato200908}),
where combinatorial arguments are quite important as well as the four or
five dimensional gauge theory.

\subsection{Orbifold projection and root of unity limit}

To construct the orbifold partition function, let us first clarify the
orbifold action for the combinatorial partition function.
Recalling the partition function is obtained from the Chern character of
the fixed point in the moduli space under the torus action, or
the equivariant $K$-theory class, it is reasonable to observe an effect of
the orbifolding on the torus action.
Their expressions are closely related to the five dimensional partition
function, so that we first study the five dimensional function
rather than the four dimensional version to discuss the meaning of the
orbifold action.

By considering ADHM construction and the localization method for
the ALE space \cite{Bianchi:1996zj,Fucito:2004ry} as in the case of the usual
$\R^4$ space, it is shown the orbifold partition function should
be defined as a sector of the $q$-deformed function (with $q \to 1$),
which is invariant under the following orbifold action
\begin{equation}
 \Gamma: \quad
 q \longrightarrow \omega q, \qquad
 t \longrightarrow \omega t, \qquad
 Q_l \longrightarrow \omega^{p_l} Q_l,
 \label{orb01}
\end{equation}
where $\omega = \exp (2\pi i/k)$ is the primitive $k^{\rm th}$ root of
unity.
$p_l$ is an integer $p_l \in \{0, \cdots, k-1\}$, and $p_{lm}=p_l-p_m$.
It denotes the $1^{\rm st}$ Chern class of the instanton bundle
\cite{Fucito:2004ry}, and also is interpreted as the linking number of the
five branes \cite{Witten:2009xu}.
Actually the boundary of the ALE space is a lens space $S^3/\Gamma$, and
thus we can assign non-trivial flat connection at infinity in this case.
Therefore the instanton solution is classified by both the $1^{\rm st}$
and $2^{\rm nd}$ Chern classes.
We remark the instanton number defined as $k/|\Gamma|$ coincides
with the $2^{\rm nd}$ Chern class only if the $1^{\rm st}$ Chern class
is vanishing.
This constraint is quite non-trivial, but it does not concern our
derivation of the matrix model discussed later.

\begin{figure}[t]
 \begin{center}
  \includegraphics[width=13em]{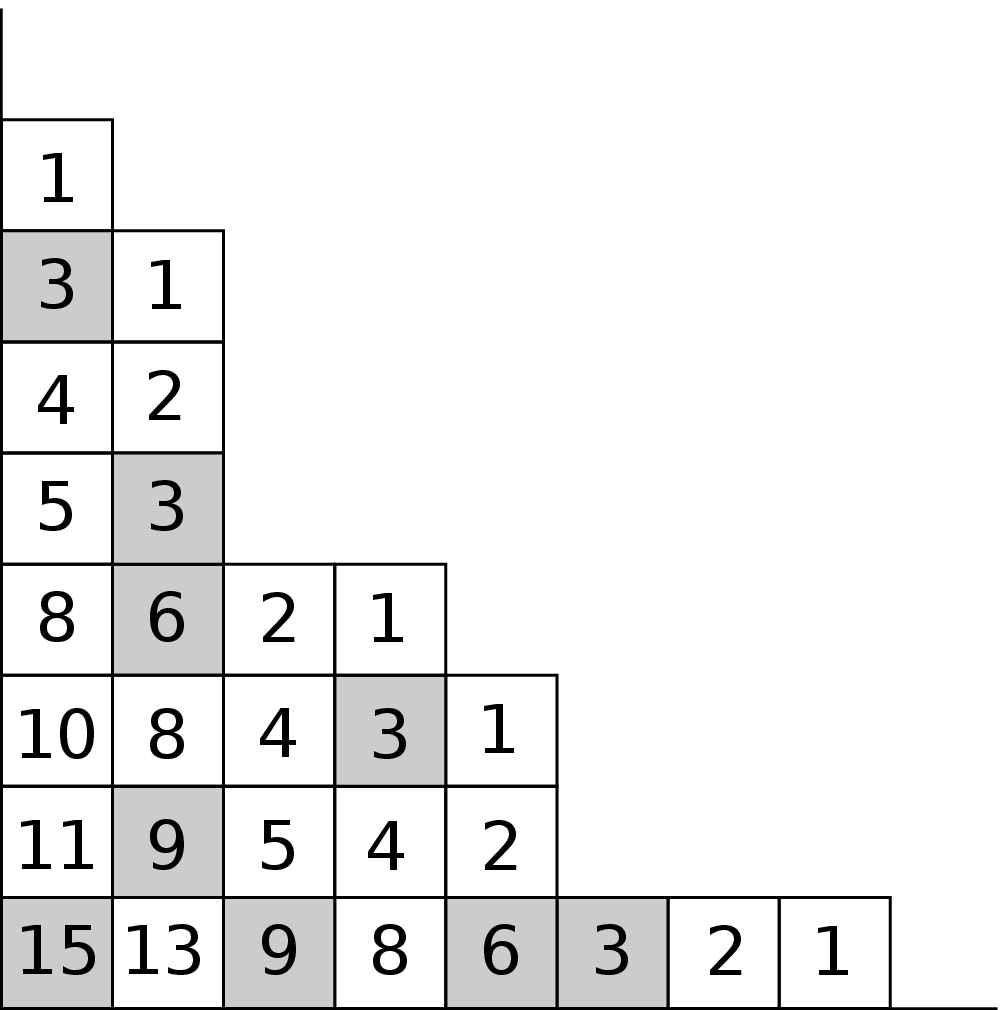}
 \end{center}
 \caption{$\Gamma$-invariant sector for $\U(1)$
 theory with $\lambda=(8,5,5,4,2,2,2,1)$. 
 Numbers in boxes stand for their hook lengths $h(i,j) = \lambda_i - j +
 \check{\lambda}_j - i + 1$.
 Shaded boxes are invariant under the action of $\Gamma=\Z_3$.
 }
 \label{part_fig01}
\end{figure}

It is easy to see this action in terms of a Young diagram.
Since the five dimensional partition function is written in the
following form
\begin{equation}
 Z_{\rm vec}^{\rm 5d} \sim
  \prod_{l,m}^n \prod_{i,j}^\infty
   \frac{1}
   {1 - Q_{lm} q^{\lambda_i^{(l)}-j} 
   t^{\check{\lambda}_j^{(m)}-i+1}},
\end{equation}
where $\check{\lambda}_j$ stands for the transposed partition, we can
see the $\Gamma$-invariant sector satisfies
\begin{equation}
 \lambda_i^{(l)}+\check{\lambda}_j^{(m)}-i-j+1+p_{lm}
  \equiv 
  0 \quad ({\rm mod}~k).
  \label{gamma-inv}
\end{equation}
For $\U(1)$ case the left hand side coincides with the hook length of
the box, $h(i,j) = \lambda_i - j + \check{\lambda}_j - i + 1$.
Thus the corresponding four dimensional function can be given by taking
into account this condition,
\begin{eqnarray}
 Z_{\rm vec}^{\rm ALE} & \sim &
  \prod_{l,m}^n
   \prod_{\Gamma\mathchar`-{\rm inv.}}
   \frac{1}
  {a_{lm} +
  \epsilon_1(i-\check{\lambda}_j^{(m)}-1)+\epsilon_2(\lambda_i^{(l)}-j)}.
  \label{part_orb01}
\end{eqnarray}
The product in this expression is explicitly taken over the
$\Gamma$-invariant sector (\ref{gamma-inv}).
Fig.~\ref{part_fig01} shows $\Gamma$-invariant sector for $\U(1)$ theory
with $\Gamma=\Z_3$.

Although this definition seems natural for the orbifold version of the
partition function, it is not useful anyway because we have to extract
the $\Gamma$-invariant sector by hands.
This procedure is called the orbifold projection in \cite{Fucito:2004ry}.
On the other hand, when we consider another deformation of the partition
function, which is given by replacing the parameters with (\ref{orb01})
as
\begin{eqnarray}
  Z_{\rm vec}^{\rm 5d} 
   & \sim &
  \prod_{l,m}^n \prod_{i,j}^\infty
   \frac{1}
   {1 - \omega^{\lambda_i^{(l)}+\check{\lambda}_j^{(m)}-i-j+1+p_{lm}} 
   Q_{lm} q^{\lambda_i^{(l)}-j} 
   t^{\check{\lambda}_j^{(m)}-i+1}},
   \label{part_orb02}
\end{eqnarray}
and take the limit of $q \to 1$, namely the root of unity limit
$q\to\exp(2\pi i/k)$ of the original $q$-partition function, 
we can see only the $\Gamma$-invariant sector contributes to the partition
function and the others are decoupled in this limit.
It is because, unless the power of $\omega$ becomes
$\lambda_i^{(l)}+\check{\lambda}_j^{(m)}-i-j+1+p_{lm} \equiv 0 ~ ({\rm
mod}~ k)$ in (\ref{part_orb02}), they just give a factor independent of
the shape of the Young diagram, or the partition. % with $q \to 1$.
Thus, if we take into account the adjoint matter contribution to regularize
the singular behavior at $q \to 1$, the weight function
behaves as
\begin{eqnarray}
 &&
  \frac
  {(1-\omega^{\lambda_i^{(l)}+\check{\lambda}_j^{(m)}-i-j+1+p_{lm}} 
  Q_{lm}Q_{\mathfrak{m}} q^{\lambda_i^{(l)}-j} t^{\check{\lambda}_j^{(m)}-i+1})}
  {(1-\omega^{\lambda_i^{(l)}+\check{\lambda}_j^{(m)}-i-j+1+p_{lm}} 
  Q_{lm} q^{\lambda_i^{(l)}-j} t^{\check{\lambda}_j^{(m)}-i+1})}
  \nonumber \\
  & \longrightarrow &
  \left\{
   \begin{array}{ccc}
    \frac
     {a_{lm} + 
     \epsilon_1(i-\check{\lambda}_j^{(m)}-1)+\epsilon_2(\lambda_i^{(l)}-j)
     + \mathfrak{m}}
     {a_{lm} +
  \epsilon_1(i-\check{\lambda}_j^{(m)}-1)+\epsilon_2(\lambda_i^{(l)}-j)}
  & \mbox{if} 
  & {\lambda_i^{(l)}+\check{\lambda}_j^{(m)}-i-j+1+p_{lm}} \equiv 0~
     (\mbox{mod}~k) \\
    1 & \mbox{if} &
     {\lambda_i^{(l)}+\check{\lambda}_j^{(m)}-i-j+1+p_{lm}} \not\equiv 0~
     (\mbox{mod}~k) \\
   \end{array}
  \right. . \nonumber \\
\end{eqnarray}
This means the orbifold projection is
automatically assigned by this parametrization.
Therefore let us define the partition function modified with
(\ref{orb01}) as the ($q$-deformed) orbifold partition function from
now.
We note that the pure Yang-Mills contribution can be extracted by taking the
decoupling limit $\mathfrak{m}\to\infty$.

We now check this reduction with a simple example, $\SU(2)$ gauge theory on
$\C^2/\Z_2$ with the adjoint matter.
If we set $\epsilon_2 = -\epsilon_1 = \hbar$, $a_1 = -a_2 = a$
and $p_1 = p_2$, lower degree parts of the instanton partition function
are obtained by replacing $q$-parameters as $q \to -q$, $Q_{12} \to
Q_{12}$ \cite{Fucito:2004ry},
\begin{eqnarray}
 Z_{\tiny\yng(1)} & = & 2
  \frac{(1+Q_{\mathfrak{m}}q)(1+Q_{\mathfrak{m}}q^{-1})}{(1+q)(1+q^{-1})}
  \frac{(1-Q_{\mathfrak{m}}Q_{12})(1-Q_{\mathfrak{m}}Q_{21})}
  {(1-Q_{12})(1-Q_{21})}, \\
 Z_{{\tiny \yng(1)},{\tiny \yng(1)}} & = & 
  \frac{(1+Q_{\mathfrak{m}}Q_{12}q)(1+Q_{\mathfrak{m}}Q_{21}q^{-1})}
       {(1+Q_{12}q)(1+Q_{21}q^{-1})}
  \frac{(1+Q_{\mathfrak{m}}Q_{12}q^{-1})(1+Q_{\mathfrak{m}}Q_{21}q)}
       {(1+Q_{12}q^{-1})(1+Q_{21}q)}
  \left(
   \frac{(1+Q_{\mathfrak{m}}q)(1+Q_{\mathfrak{m}}q^{-1})}{(1+q)(1+q^{-1})}
  \right)^2,
  \nonumber \\
 && \\
 Z_{\tiny\yng(2)} & = & 2 
  \frac{(1+Q_{\mathfrak{m}}q)(1+Q_{\mathfrak{m}}q^{-1})}{(1+q)(1+q^{-1})}
  \frac{(1-Q_{\mathfrak{m}}q^2)(1-Q_{\mathfrak{m}}q^{-2})}{(1-q^2)(1-q^{-2})}
  \nonumber \\
 && \times
  \frac{(1-Q_{\mathfrak{m}}Q_{12})(1-Q_{\mathfrak{m}}Q_{21})}
       {(1-Q_{12})(1-Q_{21})}
  \frac{(1+Q_{\mathfrak{m}}Q_{12}q)(1+Q_{\mathfrak{m}}Q_{21}q^{-1})}
       {(1+Q_{12}q)(1+Q_{21}q^{-1})}.
\end{eqnarray}
Thus, taking the limit $q\to 1$, we have
\begin{eqnarray}
 Z_{\tiny\yng(1)} & \to & 2 \left( 1 - \frac{\mathfrak{m}^2}{4a^2} \right), \\
 Z_{{\tiny \yng(1)}, {\tiny\yng(1)}} & \to & 1, \\
 Z_{\tiny\yng(2)} & \to & 2
  \left( 1 - \frac{\mathfrak{m}^2}{4\hbar^2}\right)
  \left( 1 - \frac{\mathfrak{m}^2}{4a^2}\right).
\end{eqnarray}
These correctly reproduce the result derived in \cite{Fucito:2004ry}.

Let us consider the meaning of this parametrization without taking the
four dimensional limit $|q| \to 1$.
A candidate of the corresponding manifold is the Taub-NUT manifold because
it can be obtained by compactifying the singularity on $S^1$, and
reproduces the ALE space by taking a certain limit.
It is just a speculation, thus we have to come back to this
identification problem in a future work.

We then comment on ambiguity of this parametrization.
In this paper we apply the primitive root of unity to define the
partition function as (\ref{part_orb02}), but other roots of unity,
represented as $\omega^r=\exp(2r\pi i /k)$, are
also valid for the orbifold projection while $k$ and $r$ are co-prime.
This arbitrariness reflects $\Z_k$ symmetry of the system.

\begin{figure}[t]
 \begin{center}
  \includegraphics[width=13em]{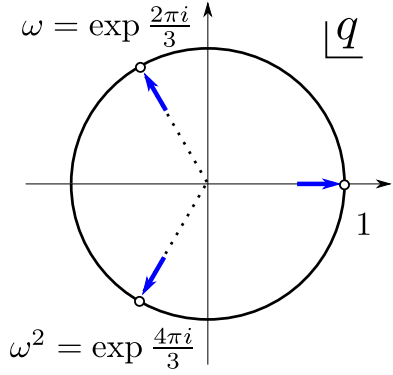}
 \end{center}
 \caption{Parametrization of $q$ corresponding to $\N=2$ theories on
 $\C^2$ and $\C^2/\Z_3$ with the limit of $|q|
 \to 1$. $|q|$ is related to the radius of the compactified dimension $S^1$.}
 \label{fig:q}
\end{figure}

Fig.~\ref{fig:q} shows how to parametrize $q$ to obtain the partition
functions for the ALE spaces.
$|q|$ corresponds to the radius of the compactified dimension $S^1$.
When we consider the four dimensional limit, we have to approach $|q|
\to 1$ with taking care of the radius of convergence since the partition
function includes infinite products of $q$.
In \cite{Fucito:2004ry} a similar parametrization is actually proposed,
which is interpreted as an analytic continuated one,
but they do not take into account regularizing the infinite
product appearing in the partition function.
Thus the parametrization used in this paper should be more suitable.

\subsection{Rearranging partitions}

As discussed before the partition function for $\SU(n)$ theory on
$\R^4$ is represented with $n$-tuple partition.
For this case, we can reproduce the one-matrix model
by blending $n$-tuple partition to a
single one \cite{Klemm:2008yu,Sulkowski:2009br,Sulkowski:2009ne}.
Next we consider its natural generalization to
the orbifold theory.
In this case, on the other hand, it is convenient for the
discussion below to divide a $n$-tuple partition into a $kn$-tuple one
\cite{Dijkgraaf:2007fe},
\begin{eqnarray}
  \left\{ 
   k(\lambda_i^{(l,r)}-i+N^{(l,r)}+p^{(l,r)})+r \Big| i = 1, \cdots, N^{(l,r)} 
  \right\}
  & = & \left\{\lambda^{(l)}_i - i + N^{(l)} + p_l \Big| i = 1, \cdots,
     N^{(l)}\right\}
  \nonumber \\
  \label{quiv_decom}
\end{eqnarray}
where
\begin{equation}
 \sum_{r=0}^{k-1} p^{(l,r)} = p_l, \qquad
 \sum_{r=0}^{k-1} N^{(l,r)} = N^{(l)}.
\end{equation}
This corresponds to the decomposition into irreducible representations
of $\Gamma$.

\begin{figure}[t]
 \begin{center}
  \includegraphics[width=35em]{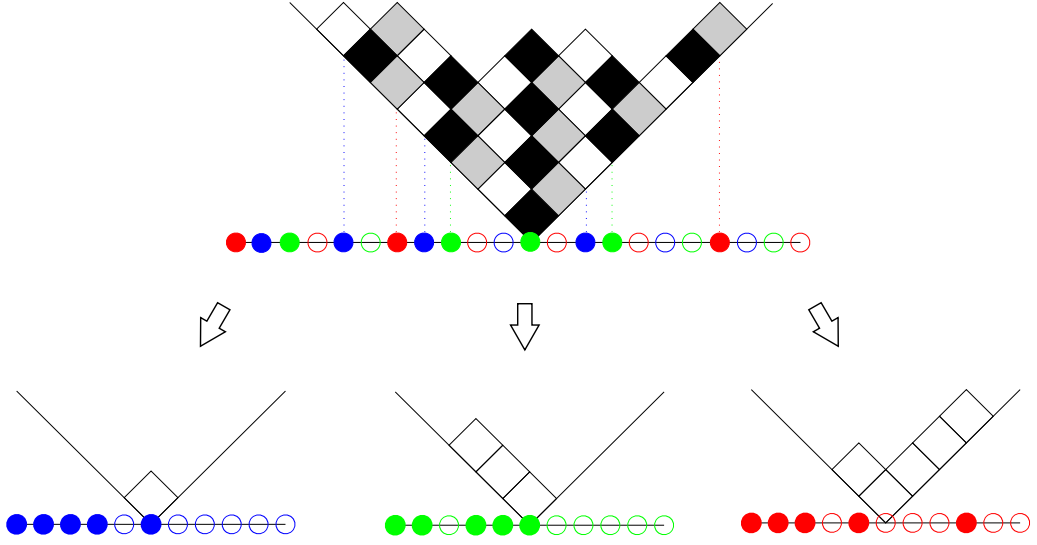}
 \end{center}
 \caption{The decomposition of the partition for $k=3$.}
 \label{part_fig02}
\end{figure}

Let us practice with an example for $\U(1)$ theory as shown in
Fig.~\ref{part_fig02}.
Starting with a partition $\lambda=(8,5,5,4,3,2,2,2,1)$, we obtain a
configuration $(15,11,10,8,5,4,3,1)$ via the mapping $\lambda_i \to
\lambda_i + N - i + p$ with $N=8$ and $p=0$.
This is interpreted as the mapping of a Young diagram to a Maya diagram,
or equivalently bosonic to fermionic variables.
Classifying entries by modulo 3, we have three configurations $(15,3)\to(5,1)$,
$(10,4,1)\to(3,1,0)$ and $(11,8,5)\to(3,2,1)$.
We then reproduce three Young diagrams from them by specifying $N^{(r)}$
and $p^{(r)}$.

%This decomposition corresponds to $\U(n)\to \U(n_0)\times \cdots \times
%\U(n_{k-1})$ as discussed in Chern-Simons matrix model on the lens space
%$S^3/\Z_k$ \cite{Aganagic:2002wv}... and also the case of the
%instantons on the ALE spaces \cite{Fucito:2004ry}.

We define the explicit relation between elements from both sides of
(\ref{quiv_decom}) as
\begin{equation}
 k(\lambda_i^{(l,r)}-i+N^{(l,r)}+p^{(l,r)})+r \equiv \lambda^{(l)}_j - j
  +N^{(l)}+ p_l
  \quad \mbox{with} \quad j = c^{(l,r)}_i,
\end{equation}
Here $c^{(l,r)}_i$ means the mapping from the index of the divided
$kn$-partition to that of the original $n$-partition.
Introducing another set of variables
\begin{equation}
 h_i^{(l,r)} \equiv k(\lambda_i^{(l,r)}-i+N^{(l,r)}+p^{(l,r)}), \qquad
 h_i^{(l)} \equiv \lambda_i^{(l)}-i+N^{(l)}+p_l, 
\end{equation}
\begin{equation}
 \ell_i^{(l,r)} \equiv h_i^{(l,r)} + b_l - p_l + r, 
\end{equation}
and setting $N^{(l)}=\bf{N}$ and
$N^{(l,r)}=N$ for all $l=1, \cdots, n$ and $(l,r)=(1,0), \cdots, (n,k-1)$
for simplicity,
the partition function (\ref{part_vec02}) is, up to constants,
rewritten as,
\begin{eqnarray}
 Z_{\rm vec} & = & \prod_{(l,i)\not=(m,j)}
 \frac{(\omega^{h_i^{(l)}-h_j^{(j)}}
        q^{h_i^{(l)}-h_j^{(j)}+(\beta-1)(j-i)+b_{lm}-p_{lm}};\tilde q)_\infty}
      {(\omega^{h_i^{(l)}-h_j^{(j)}+1}
        q^{h_i^{(l)}-h_j^{(j)}+(\beta-1)(j-i)+b_{lm}-p_{lm}+\beta};\tilde q)_\infty}
 \nonumber \\
 && \times \prod_{l,m}^n \prod_{i=1}^{\bf N}
  \frac{(\omega^{h_i^{(l)}-p_l+1}
         q^{h_i^{(l)}+(\beta-1)({\bf N}-i)+b_{lm}-p_l+\beta};\tilde q)_\infty}
       {(\omega^{-h_i^{(l)}+p_l}
         q^{-h_i^{(l)}-(\beta-1)({\bf N}-i)-b_{lm}+p_l};\tilde q)_\infty}
 \nonumber \\
 & = & \prod_{(l,r,i)\not=(m,s,j)}
  \frac{(\omega^{r-s} q^{\ell^{(l,r)}_i-\ell^{(m,s)}_j+(\beta-1)
        (c^{(m,s)}_j-c^{(l,r)}_i)};\tilde q)_\infty}
       {(\omega^{r-s+1} q^{\ell^{(l,r)}_i-\ell^{(m,s)}_j+(\beta-1)
        (c^{(m,s)}_j-c^{(l,r)}_i)+\beta};\tilde q)_\infty}
 \nonumber \\
 & & \times \prod_{l,m}^n \prod_{r=0}^{k-1} \prod_{i=1}^{N}
    \frac{(\omega^{r-p_l+1} q^{\ell^{(l,r)}_i-b_m+(\beta-1)
          ({\bf N}-c^{(l,r)}_i)+\beta};\tilde q)_\infty}
         {(\omega^{-r+p_l} q^{-(\ell^{(l,r)}_i-b_m+(\beta-1)
          ({\bf N}-c^{(l,r)}_i))};\tilde q)_\infty}.	 
\end{eqnarray}
The former expression is in terms of the original $n$-tuple
representation, and the latter is written with a $kn$-tuple partition.
Blending the $kn$-tuple to $k$-tuple partitions as
\begin{equation}
 \ell_{i=1, \cdots, \sum_{l=1}^n N^{(l,r)}}^{(r)}
  = \left(
     \ell_{1}^{(n,r)}, \cdots, \ell_{N^{(n,r)}}^{(n,r)}, \cdots,
     \ell_{1}^{(1,r)}, \cdots, \ell_{N^{(1,r)}}^{(1,r)}
    \right),
\end{equation}
we finally obtain an expression in terms of $k$-tuple partition,
\begin{eqnarray}
  Z_{\rm vec} & = &
 \prod_{(r,i)\not=(s,j)}
  \frac{(\omega^{r-s} q^{\ell^{(r)}_i-\ell^{(s)}_j+(\beta-1)
        (c^{(s)}_j-c^{(r)}_i)};\tilde q)_\infty}
       {(\omega^{r-s+1} q^{\ell^{(r)}_i-\ell^{(s)}_j+(\beta-1)
        (c^{(s)}_j-c^{(r)}_i)+\beta};\tilde q)_\infty}
 \nonumber \\ && \times	
 \prod_{l=1}^n \prod_{r=0}^{k-1} \prod_{i=1}^{nN}
    \frac{(\omega^{r-p_l+1} q^{\ell^{(r)}_i-b_l+(\beta-1)
          ({\bf N}-c^{(r)}_i)+\beta};\tilde q)_\infty}
         {(\omega^{-r+p_l} q^{-(\ell^{(r)}_i-b_l+(\beta-1)
          ({\bf N}-c^{(r)}_i))};\tilde q)_\infty}.
 \label{part_vec03}
\end{eqnarray}
Again $c^{(r)}_i$ stands for the mapping from the index of the $k$-tuple
partition to that of $n$-tuple one.
We will discuss its matrix model description from the expression of
(\ref{part_vec03}) in section \ref{sec:matrix}.

We can also get explicit representations for matter parts in a similar way,
\begin{eqnarray}
  Z_{\rm adj} & = &
 \prod_{(r,i)\not=(s,j)}
  \frac{(\omega^{r-s+1} q^{\ell^{(r)}_i-\ell^{(s)}_j+\mathfrak{M}+(\beta-1)
        (c^{(s)}_j-c^{(r)}_i)+\beta};\tilde q)_\infty}
       {(\omega^{r-s} q^{\ell^{(r)}_i-\ell^{(s)}_j+\mathfrak{M}+(\beta-1)
        (c^{(s)}_j-c^{(r)}_i)};\tilde q)_\infty}
 \nonumber \\	
 && \times
  \prod_{l=1}^n \prod_{r=0}^{k-1} \prod_{i=1}^{nN}
    \frac{(\omega^{-r+p_l} q^{-(\ell^{(r)}_i-b_l+\mathfrak{M}+(\beta-1)
          ({\bf N}-c^{(r)}_i))};\tilde q)_\infty}
         {(\omega^{r-p_l+1} q^{\ell^{(r)}_i-b_l+\mathfrak{M}+(\beta-1)
          ({\bf N}-c^{(r)}_i)+\beta};\tilde q)_\infty},         
 \label{part_adj02}
\end{eqnarray}
\begin{eqnarray}
 Z_{\rm fund} & = & 
  \prod_{l=1}^n \prod_{f=1}^{N_f} \prod_{r=0}^{k-1} \prod_{i=1}^{nN}
  \frac{(\omega^{-i+p_l+1}q^{b_l+M_f-\beta c_i^{(r)}+1};\tilde q)_\infty}
       {(\omega^{r-N+1}q^{\ell_i^{(r)}+M_f-(\beta-1)c_i^{(r)}-{\bf N}+1};\tilde q)_\infty}
       , \label{part_fund05} \\
 Z_{\rm antifund} & = & 
  \prod_{l=1}^n \prod_{f=1}^{N_f} \prod_{r=0}^{k-1} \prod_{i=1}^{nN}
  \frac{(\omega^{-i+p_l+1}q^{b_l-M_f-\beta c_i^{(r)}+\beta};\tilde q)_\infty}
       {(\omega^{r-N+1}q^{\ell_i^{(r)}-M_f-(\beta-1)c_i^{(r)}-{\bf N}+\beta};\tilde q)_\infty}.
       \label{part_fund06}
\end{eqnarray}

\section{Matrix model description}\label{sec:matrix}

We then derive matrix models from the orbifold partition
functions.
According to the result that the orbifold partition function can be
written with a $k$-tuple partition, one can see the $k$-matrix model
is naturally arising from the combinatorial expression.
The multi-matrix model, derived in this section, is
\begin{equation}
 Z = \int \mathcal{D}\vec{X}
  e^{-\frac{1}{\epsilon_2}\sum_{r=0}^{k-1}\sum_{i=1}^{N} V(x_i^{(r)})}
\end{equation}
\begin{equation}
  \mathcal{D}\vec{X} = 
  \prod_{r=0}^{k-1} \prod_{i=1}^{N} \frac{d x_i^{(r)}}{2\pi}
  \Delta^2(x), 
\end{equation}
\begin{equation}
 V(x) = V_{\rm vec}(x) + V_{\rm (anti)fund}(x).
\end{equation}
We remark this matrix model corresponds to only the contribution from the
instanton part, thus we have to introduce the perturbative piece when we
consider the whole prepotential of the gauge theory.
%The linear term $\tau x$ corresponds to the dynamical scale in the
%partition function, which is also interpreted as the counting parameter.
The deformed version of the Vandermonde determinant appearing in the matrix
measure is given by
\begin{eqnarray}
 \Delta^2(x) & = &
  \prod_{r=0}^{k-1} \prod_{i<j}^{N}
  \left(
   2 \sinh \frac{x_i^{(r)}-x_j^{(r)}}{2}
  \right)^2
  \prod_{r<s}^{k-1} \prod_{i,j}^{N}
  \left(
   2 \sinh \frac{x_i^{(r)}-x_j^{(s)}+d^{(r,s)}}{2}
  \right)^2
  \nonumber \\
 && \times
  \prod_{r=0}^{k-1} \prod_{i<j}^{N}
  \left(
   2 \sinh \frac{k(x_i^{(r)}-x_j^{(r)})}{2}
  \right)^{2\gamma}
  \prod_{r<s}^{k-1} \prod_{i,j}^{N}
  \left(
   2 \sinh \frac{k(x_i^{(r)}-x_j^{(s)})}{2}
  \right)^{2\gamma}
 \label{vander01}
\end{eqnarray}
with 
\begin{equation}
 d^{(r,s)} = \frac{2\pi i}{k} (r-s).
\end{equation}
Here we introduce another parameter $\gamma$, which is related to $\beta$ as
$\beta=k\gamma+1$.
Note that $N$ stands for the matrix size here.
This matrix measure coincides with that of the Chern-Simons matrix model on the
lens space $S^3/\Z_k$
\cite{Aganagic:2002wv,Halmagyi:2003ze,Halmagyi:2003mm} when we take
$\gamma=0$, corresponding to $\beta=1$.
% since it is defined as (\ref{Ug_param}).
In section \ref{sec:spectral} we will study this case in detail.

To obtain a matrix model from the partition function, we consider
asymptotic behavior of the combinatorial expressions by introducing the
following variables 
\begin{equation}
 x_i^{(r)} = \frac{\ell_i^{(r)}}{\epsilon_2},
\end{equation}
and taking the limit of $\epsilon_2 \to 0$.
We study the asymptotics of the orbifold partition functions in the
following.

\subsection{Matrix measure}

Let us start with the measure part of the multi-matrix model, coming
 from the combinatorial expression of (\ref{part_vec03}).
In this derivation we assume
\begin{equation}
 \beta = k \gamma + 1 \equiv 1 \quad ({\rm mod}~k),
  \label{Ug_param}
\end{equation}
in order to satisfy the condition
\begin{equation}
 \omega q^\beta = (\omega q)^\beta.
  \label{Ug_lim}
\end{equation}
This parametrization is proposed in \cite{Uglov:1997ia}, thus we call
it the {\it Uglov condition}.\footnote{
The reason why we assign this condition (\ref{Ug_param}) here is just a
technical one.
We will see, in the forthcoming paper \cite{Kimura:2011gq}, that this
condition is not essential for deriving the matrix model description.
}
One can see it is quite important to obtain the matrix model
description from the combinatorial expression because if we denote
$\tilde q = \omega q$, this condition implies
\begin{equation}
 (q,t) \longrightarrow (\omega q, \omega t) = (\tilde q, \tilde q^\beta).
\end{equation}

Due to the condition (\ref{Ug_lim}), the infinite product in
(\ref{part_vec03}) is reduced as
\begin{eqnarray}
 &&
  \prod_{(r,i)\not=(s,j)}
  \frac{(\omega^{r-s} q^{\ell^{(r)}_i-\ell^{(s)}_j+(\beta-1)
        (c^{(s)}_j-c^{(r)}_i)};\tilde q)_\infty}
       {(\omega^{r-s+1} q^{\ell^{(r)}_i-\ell^{(s)}_j+(\beta-1)
        (c^{(s)}_j-c^{(r)}_i)+\beta};\tilde q)_\infty}
 \nonumber \\
 & = & \prod_{(r,i)\not=(s,j)} \prod_{p=0}^{\beta-1}
  \left(
   1 - \omega^{r-s+p} q^{\ell^{(r)}_i-\ell^{(s)}_j+(\beta-1)
        (c^{(s)}_j-c^{(r)}_i)+p}
  \right).
\end{eqnarray}
Taking the limit $\epsilon_2 \to 0$ and using (\ref{Ug_param}), we can
treat them as
\begin{equation}
 \left(
  1 - \omega^{r-s} e^{x^{(r)}_i-x^{(s)}_j}
 \right)
 \prod_{p=1}^{k\gamma}
 \left(
  1 - \omega^{r-s+p} e^{x^{(r)}_i-x^{(s)}_j}
 \right)
 = 
 \left(
  1 - \omega^{r-s} e^{x^{(r)}_i-x^{(s)}_j}
 \right)
 \left(
  1 - e^{k(x^{(r)}_i-x^{(s)}_j)}
 \right)^\gamma
\end{equation}
where we use the identity
\begin{equation}
 \prod_{r=0}^{k-1} (1-\omega^r z) = (1-z^k).
\end{equation}
As a result we obtain the deformed Vandermonde determinant $\Delta^2(x)$
presented in (\ref{vander01}).

When we take the four dimensional limit, a factor including $d^{(r,s)}$
does not contribute in the leading order because it is expanded as
\begin{eqnarray}
 2 \sinh \left(\frac{x_i^{(r)}-x_j^{(s)}+d^{(r,s)}}{2}\right)
  = 2 i \sin \left(\frac{r-s}{k}\pi\right)
  \left[
   1 - i \left(\frac{x_i^{(r)}-x_j^{(r)}}{2}\right) 
   \cot \left(\frac{r-s}{k}\pi\right) 
   + \cdots
  \right]. \nonumber \\
\end{eqnarray}
Thus the matrix measure goes to
\begin{equation}
 \Delta^2(x) \longrightarrow
 \prod_{r=0}^{k-1} \prod_{i<j}^{nN}
  \left(
   x_i^{(r)}-x_j^{(r)}
  \right)^{2+2\gamma}
 \prod_{r<s}^{k-1} \prod_{i,j}^{nN}
  \left(
   x_i^{(r)}-x_j^{(s)}
  \right)^{2\gamma}.
  \nonumber \\
 \label{vander4d}
\end{equation}
This shows there is no interaction between $k$ matrices in this part
when we take $\beta=1$ $(\gamma=0)$, corresponding to the most desirable
$\SU(n)$ theory.
We remark that it can be apparently written in a simple form in terms of
$k$-matrix model, but it is still difficult to write it down with the
original combinatorial representation.

\subsection{Matrix potentials}

We then discuss the potential terms for the matrix model.
To obtain the matrix potentials, we have to evaluate the
asymptotic behavior of the quantum dilogarithm function
\cite{Eynard:2008mt}\cite{Klemm:2008yu,Sulkowski:2009br,Sulkowski:2009ne}
\begin{equation}
 g(z;q) = \prod_{p=1}^\infty 
  \left(
   1 - \frac{1}{z} q^p
  \right).
  \label{qdilog01}
\end{equation}
While the series expansion with
$q=e^{\epsilon_2}$ has been already investigated, an explicit expression
for expansion around the root of unity, which is required for studying the
orbifold partition function, is not revealed yet.
Substituting $\tilde q = \omega e^{\epsilon_2}$ into (\ref{qdilog01}),
we obtain a similar expression
\begin{eqnarray}
 \log g(z;\tilde q) & = &
 \frac{1}{k \epsilon_2} \sum_{r=0}^{k-1} \sum_{n=0}^\infty
 {\rm Li}_{2-n}\left(\frac{\omega^r}{z}\right) (- \epsilon_2)^n
 \sum_{m=0}^n \frac{(k-r)^{n-m}k^m}{(n-m)!m!} B_m
 \nonumber \\
 & = & \frac{1}{\epsilon_2}
  \left[
   \frac{1}{k^2} {\rm Li}_2\left(\frac{1}{z^k}\right)
   + \mathcal{O}(\epsilon_2)
  \right].
\end{eqnarray}
Here $B_n$ stands for the $n^{\rm th}$ Bernoulli number and ${\rm
Li}_n(z)$ is the polylogarithm function
\begin{equation}
 {\rm Li}_n(z) = \sum_{p=1}^\infty \frac{z^p}{p^n}.
  \label{polylog01}
\end{equation}
For $n=2$ it is especially called the {\it dilogarithm}.
Since the expression of (\ref{polylog01}) is valid where $|z|<1$, it is
convenient to introduce the useful identity, called the inversion formula,
\begin{equation}
 {\rm Li}_2 (z) + {\rm Li}_2 (1/z) =
  - \frac{1}{2} (\log z)^2 + \frac{\pi^2}{3} 
  - i \pi \log z.
  \label{polylog02}
\end{equation}
Thus we can extend its domain to the whole complex plane.

Utilizing the results obtained above, we can evaluate the factor
contributing to the matrix potential.
The vector multiplet part in (\ref{part_vec03}) yields
\begin{eqnarray}
 \prod_{l=1}^n \prod_{r=0}^{k-1} \prod_{i=1}^{nN}
    \frac{(\omega^{r-p_l+1} q^{\ell^{(r)}_i-b_l+(\beta-1)
          ({\bf N}-c^{(r)}_i)+\beta};\tilde q)_\infty}
         {(\omega^{-r+p_l} q^{-(\ell^{(r)}_i-b_l+(\beta-1)
          ({\bf N}-c^{(r)}_i))};\tilde q)_\infty}
 & \equiv & \exp \sum_{r=0}^{k-1} \sum_{i=1}^{nN} 
  - \frac{1}{\epsilon_2} V^{\rm 5d}_{\rm vec}(x_i^{(r)}),
  \nonumber \\
\end{eqnarray}
\begin{eqnarray}
 V^{\rm 5d}_{\rm vec}(x) & = & - \frac{1}{k^2} \sum_{l=1}^n
  \left[
   {\rm Li}_2(e^{k(x-a_l)}) - {\rm Li}_2(e^{-k(x-a_l)})
  \right]
  + \mathcal{O}(\epsilon_2)
  \nonumber \\
 & \simeq & \frac{n}{2} x^2 + \frac{2}{k^2} \sum_{l=1}^n {\rm
  Li}_2(e^{-k(x-a_l)}).
\end{eqnarray}
Here we neglect a redundant constant term in this potential, and if
$x-a_l<0$, we have to redefine this by using the identity
(\ref{polylog02}).
The derivative of this potential is given by
\begin{equation}
  {V^{\rm 5d}_{\rm vec}}'(x) = 
   \frac{2}{k} \sum_{l=1}^n \log 
   \left[ 2 \sinh \left(\frac{k}{2} (x-a_l) \right)
   \right].
\end{equation}
We can also derive the potential for the four dimensional theory.
Taking the four dimensional limit, we have
\begin{equation}
 V^{\rm 4d}_{\rm vec}(x) = \frac{2}{k} \sum_{l=1}^n 
  \left[
   (x-a_l) \log (x-a_l) - (x-a_l)
  \right],
  \label{pot_vec4d}
\end{equation}
and its derivative
\begin{equation}
 {V^{\rm 4d}_{\rm vec}}'(x) = \frac{2}{k} \sum_{l=1}^{n}
  \log (x-a_l).
\end{equation}
We will show this potential plays an important role in obtaining
Seiberg-Witten curve as the spectral curve of the matrix model.

%\subsection{Potential for a fundamental matter}

We then investigate the matrix model potentials for the fundamental
matter from the expression (\ref{part_fund05}) in a similar way,
\begin{eqnarray}
 Z_{\rm fund} & = & \prod_{r=0}^{k-1} \prod_{f=1}^{N_f}
  \prod_{i=1}^{nN}
  \frac{g(\omega^{-1+r}q^{-(M_f+1-\beta i)};\tilde q)}
       {g(\omega^{-1+r}q^{-(\ell_i^{(r)}+M_f+(\beta-1)i+1)};\tilde q)}
 \equiv \exp \sum_{r=0}^{k-1} \sum_{i=1}^{nN_0} - \frac{1}{\epsilon_2}
  V_{\rm fund}^{\rm 5d}(x_i^{(r)}),
  \nonumber \\
\end{eqnarray}
\begin{eqnarray}
 V_{\rm fund}^{\rm 5d} & = & \frac{1}{k^2} \sum_{f=1}^{N_f}
  {\rm Li}_2 (e^{k(x+m_f)}),
 \\
 V_{\rm fund}^{\rm 4d} & = & - \frac{1}{k} \sum_{f=1}^{N_f}
  \left[
   (x+m_f) \log (x+m_f) - (x+m_f)
  \right].
\end{eqnarray}
We again neglect a finite constant independent of $x$ for them.
For $k=1$ these potential functions coincide with the usual matrix model
potentials for $\N=2$ theories
\cite{Klemm:2008yu,Sulkowski:2009ne}.
We now remark this matrix potential is independent of the deformation
parameter $\beta$, and its dependence only appears in the matrix
measure.

\subsection{$\N=2^*$ theory}

Let us derive another matrix model by taking into account
the contribution from the adjoint matter, which should describe
$\N=2^*$ theory \cite{Sulkowski:2009br}.
After almost the same procedure discussed above, we
now obtain the following matrix model:
\begin{equation}
 Z = \int \mathcal{D}\vec{X}
  e^{-\frac{1}{\epsilon_2}\sum_{r=0}^{k-1}\sum_{i=1}^{N} V(x_i^{(r)})},
\end{equation}
\begin{equation}
  \mathcal{D}\vec{X} = 
  \prod_{r=0}^{k-1} \prod_{i=1}^{N} \frac{d x_i^{(r)}}{2\pi}
  \frac{\Delta^2(x)}{\Delta_{\mathfrak{m}}^2(x)}, 
\end{equation}
\begin{equation}
 V(x) = V_{\rm vec}(x) + V_{\rm adj}(x).
\end{equation}
Again $N$ stands for the matrix size here.
In this case the matrix measure requires another contribution,
\begin{eqnarray}
 \Delta_{\mathfrak{m}}^2(x) & = &
  \prod_{r=0}^{k-1} \prod_{i\not=j}^{N}
  \left(
   2 \sinh \frac{x_i^{(r)}-x_j^{(r)}+\mathfrak{m}}{2}
  \right)
  \prod_{r\not=s}^{k-1} \prod_{i,j}^{N}
  \left(
   2 \sinh \frac{x_i^{(r)}-x_j^{(s)}+d^{(r,s)}+\mathfrak{m}}{2}
  \right)
  \nonumber \\
 && \times
  \prod_{r=0}^{k-1} \prod_{i\not=j}^{N}
  \left(
   2 \sinh \frac{k(x_i^{(r)}-x_j^{(r)}+\mathfrak{m})}{2}
  \right)^{\gamma}
  \prod_{r\not=s}^{k-1} \prod_{i,j}^{N}
  \left(
   2 \sinh \frac{k(x_i^{(r)}-x_j^{(s)}+\mathfrak{m})}{2}
  \right)^{\gamma}.
  \nonumber \\
\end{eqnarray}
The matrix potential for the adjoint matter is
\begin{equation}
 V_{\rm adj}(x) = 
  - \frac{n}{2}(x+\mathfrak{m})^2
  - \frac{2}{k^2} \sum_{l=1}^n {\rm Li}_2
  \left( e^{-k(x-a_l+\mathfrak{m})} \right).
\end{equation}
As a result, the total potential function becomes
\begin{equation}
 V_{\rm vec}(x) + V_{\rm adj}(x) 
  = - n \mathfrak{m} x + 
  \frac{2}{k^2} \sum_{l=1}^n
  \left[
   {\rm Li}_2 \left( e^{-k(x-a_l)} \right)
   - {\rm Li}_2 \left( e^{-k(x-a_l+\mathfrak{m})} \right)
  \right].
\end{equation}
This result is consistent with \cite{Sulkowski:2009br} for $k=1$.

We now comment on a relation to $\N=4$ theory on the ALE spaces
\cite{Dijkgraaf:2007fe}.
Since $\N=4$ theory is obtained from $\N=2^*$ theory by
taking the massless limit of the adjoint matter $\mathfrak{m} \to 0$,
the result of \cite{Dijkgraaf:2007fe} should be related to the
result discussed above.
In the massless limit the matrix measure and the potential become
trivial
\begin{equation}
 \frac{\Delta^2(x)}{{\Delta^2_{\mathfrak{m}}(x)}} \longrightarrow 1,
  \qquad
 V(x) \longrightarrow 0.
\end{equation}
This means, when we go back to the combinatorial representation, the
combinatorial weight, represented in terms of the hook length, becomes
trivial in the partition function.
As a result it remains only the counting parameter, which is related to
the dynamical scale.

For $\U(1)$ theory, therefore it simply corresponds to the following
partition function only with implementing the orbifold projection,
\begin{equation}
 Z \sim \sum_{\lambda} \Lambda^{\#\{ 
  \mbox{\scriptsize $\Gamma$-invariant boxes} \}}.
\end{equation}
This is similar to the partition function discussed in
\cite{Dijkgraaf:2007fe}, but not identical.
We should investigate this type of partition function and clarify the
relation to $\N=4$ theory in a future work.

\section{Spectral curve of the matrix model}\label{sec:spectral}

To study a connection between the gauge theory and the matrix model, in
this section we consider the multi-matrix model in detail, which is
defined as
\begin{equation}
 Z = \int \prod_{r=0}^{k-1} \prod_{i=1}^N 
  \frac{d x_i^{(r)}}{2\pi} \Delta^2(x)
  e^{-\frac{1}{g_s}\sum_{r=0}^{k-1} \sum_{i=1}^N V(x_i^{(r)})}
\end{equation}
where $\Delta^2(x)$ is defined in (\ref{vander01}).
In this section $\epsilon_2$ is replaced with $g_s$, and we focus on the
case of $\beta=1$.
For a while we consider a generic potential $V(x)$.
The method used in this section is partially based on that
developed for the lens space Chern-Simons matrix model
\cite{Halmagyi:2003ze,Marino:2011nm}.

\subsection{Large $N$ limit and saddle point equation}

We are interested in the 't Hooft limit of this matrix model, in which
\begin{equation}
 g_s \longrightarrow 0, \qquad
 N \longrightarrow \infty,
\end{equation}
with fixing the 't Hooft coupling 
\begin{equation}
 T = g_s N.
\end{equation}
Actually, in this large $N$ limit, the evaluation of the matrix integral
reduces to the calculation of the critical points.

If we define the prepotential for the matrix model as
\begin{eqnarray}
 - \frac{1}{g_s^2} \mathcal{F}
  & = & - \frac{1}{g_s} \sum_{r=0}^{k-1} \sum_{i=1}^N V(x_i^{(r)})
  \nonumber \\
  && + 2 \sum_{r=0}^{k-1} \sum_{i<j}^N \log \sinh
  \left(
   \frac{x_i^{(r)} - x_j^{(r)}}{2}
  \right)
  \nonumber \\
  && + 2 \sum_{r<s}^{k-1} \sum_{i,j}^N \log \sinh
  \left(
   \frac{x_i^{(r)} - x_j^{(r)}+d^{(r,s)}}{2}
  \right),
\end{eqnarray}
which corresponds to the genus zero part, we would extract the information
about the gauge theory, e.g. Seiberg-Witten curve, from this function.
We can obtain the condition for criticality by differentiating the
prepotential,
\begin{equation}
 \frac{1}{g_s} V'(x_i^{(r)}) = 
  \sum_{j(\not=i)}^N \coth 
  \left(
   \frac{x_i^{(r)}-x_j^{(r)}}{2}
  \right)
  + \sum_{s(\not=r)}^{k-1} \sum_{j=1}^N \coth 
  \left(
   \frac{x_i^{(r)}-x_j^{(s)}+d^{(r,s)}}{2}
  \right).
  \label{saddle01}
\end{equation}
This saddle point equation is also given by the extremal condition for
the effective potential defined as
\begin{equation}
 V^{(r)}_{\rm eff}(x_i^{(r)})
  = V(x_i^{(r)})
  - \frac{2T}{N} \sum_{j(\not=i)}^N
  \sinh \left( \frac{x_i^{(r)}-x_j^{(r)}}{2} \right)
  - \frac{2T}{N} \sum_{s(\not=r)}^{k-1} \sum_{j=1}^N
  \sinh \left( \frac{x_i^{(r)}-x_j^{(s)}+d^{(r,s)}}{2} \right) .
\end{equation}
This potential involves a logarithmic Coulomb repulsion between
eigenvalues.
If the 't Hooft coupling is small, the potential term dominates the
Coulomb interaction and eigenvalues concentrate on extrema of the
potential $V'(x)=0$.
On the other hand, as the coupling gets bigger, the eigenvalue
distribution is extended.

To deal with this situation, we now define the densities of eigenvalues
for each matrix,
\begin{equation}
 \rho^{(r)}(x) = \frac{1}{N} 
   \sum_{i=1}^N \delta(x-x_i^{(r)})
\end{equation}
where $x_i^{(r)}$ is the solution of the criticality condition
(\ref{saddle01}).
In the large $N$ limit, it is natural to think these eigenvalue
distributions are smeared, and become continuous functions.
Furthermore, as in the case of the usual matrix model, we assume the
eigenvalues are distributed around the critical points of the potential
$V(x)$ as linear segments.
In particular there are $n$ critical points for $\SU(n)$ theory.
Thus we generically denote the $l^{\rm th}$ segment for $\rho^{(r)}(x)$
as $\mathcal{C}^{(l,r)}$, and the total number of eigenvalues $N$ splits
into $n$ integers for these segments,
\begin{equation}
 N = \sum_{l=1}^n N^{(l,r)}
\end{equation}
where $N^{(l,r)}$ is the number of eigenvalues in the interval
$\mathcal{C}^{(l,r)}$.
The density of eigenvalues $\rho^{(r)}(x)$ takes non-zero value only on
the segment $\mathcal{C}^{(l,r)}$, and is normalized as
\begin{equation}
 \int_{\mathcal{C}^{(l,r)}} dx \rho^{(r)}(x) =
  \frac{N^{(l,r)}}{N} \equiv \nu^{(l,r)}
  \label{normal}
\end{equation}
where we call it the {\it filling fraction}.
According to these fractions, we can introduce the partial 't Hooft
parameters,
\begin{equation}
 T^{(l,r)} = g_s N^{(l,r)}.
\end{equation}
Note there are $n$ 't Hooft couplings and filling fractions, but only
$n-1$ fractions are independent since they have to satisfy $\sum_{l=1}^n
\nu^{(l,r)}=1$ while all the 't Hooft couplings are independent.
We will show they are related to Coulomb moduli parameters as in the case
discussed in \cite{Dijkgraaf:2002fc,Dijkgraaf:2009pc}.

Using these distribution functions, the prepotential and the criticality
condition can be written as
\begin{eqnarray}
 \mathcal{F} & = & T \sum_{r=0}^{k-1} \int dx \rho^{(r)}(x) V(x)
  \nonumber \\ &&
  - T^2 \sum_{r=0}^{k-1} \PV \int dx dy \rho^{(r)}(x) \rho^{(r)}(y)
  \log \sinh \left( \frac{x-y}{2} \right)
  \nonumber \\ &&
  - T^2 \sum_{r\not=s}^{k-1} \int dx dy \rho^{(r)}(x) \rho^{(s)}(y)
  \log \sinh \left( \frac{x-y+d^{(r,s)}}{2} \right),
 \\
 \frac{1}{T} V'(x) & = & 
  \PV \int dy \rho^{(r)}(y) \coth \left(\frac{x-y}{2}\right)
  \nonumber \\
  && + \sum_{s(\not=r)}^{k-1} \int dy \rho^{(s)}(y) 
  \coth \left(\frac{x-y+d^{(r,s)}}{2}\right).
\end{eqnarray}
Here $\PV$ stands for the principal value.
To assign the normalization conditions (\ref{normal}), we then introduce
the Lagrange multipliers to the original prepotential,
\begin{equation}
 \mathcal{F} \longrightarrow \mathcal{F}
  + \sum_{r=0}^{k-1} \sum_{l=1}^n \Gamma^{(l,r)}
  \left( T \int_{\mathcal{C}^{(l,r)}} dx \rho^{(r)}(x) - T^{(l,r)}  \right).
 \label{legendre_part}
\end{equation}
This is just the Legendre transformation: $\Gamma^{(l,r)}$ is conjugate
to $T^{(l,r)}$.
From the variation with respect to the density of states, we obtain
\begin{eqnarray}
 V(x) & = & 2 T \int dy \rho^{(r)}(y) 
  \log \sinh \left(\frac{x-y}{2}\right)
  \nonumber \\
 &&
  + 2 T \sum_{r=0}^{k-1} \int dy \rho^{(s)}(y) 
  \log \sinh \left(\frac{x-y+d^{(r,s)}}{2}\right)
  + \Gamma^{(l,r)}, 
  \qquad x \in \mathcal{C}^{(l,r)},
  \nonumber \\
\end{eqnarray}
which is equivalent to
\begin{equation}
 V^{(r)}_{\rm eff}(x) = \Gamma^{(l,r)}, \qquad x \in \mathcal{C}^{(l,r)}.
\end{equation}

\begin{figure}[t]
 \begin{center}
  \includegraphics[width=42em]{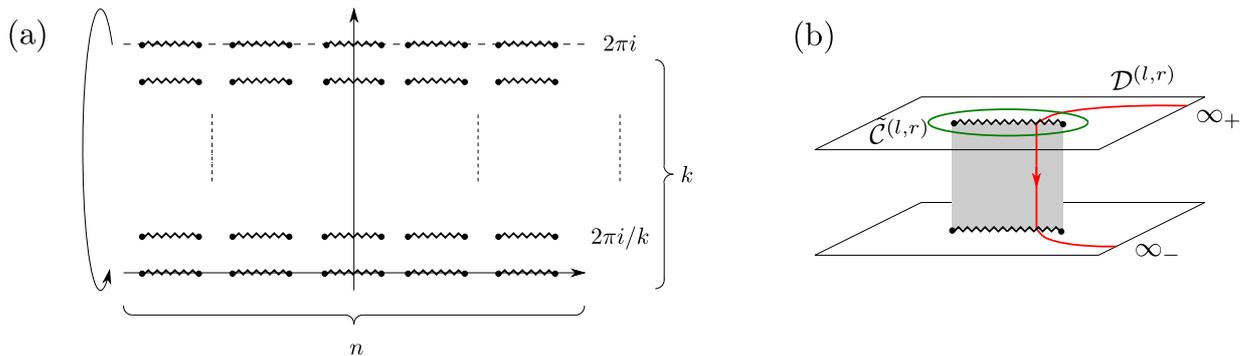}
 \end{center}
 \caption{(a) Cuts of the total resolvent on
 the complex plane. It is cylindrical since there is a periodicity in
 the imaginary direction $x \sim x + 2\pi i$. There are totally $kn$
 cuts on the plane.
 (b) Expansion around a specific cut. The cycle
 $\mathcal{D}^{(l,r)}$ is conjugate to $\tilde{\mathcal{C}}^{(l,r)}$. }
 \label{cuts}
\end{figure}

In order to solve these conditions, we then introduce the individual
resolvents
\begin{equation}
 \omega^{(r)}(x) = g_s \sum_{i=1}^N 
  \coth \left( \frac{x-x_i^{(r)}}{2} \right).
\end{equation}
Since all the singularities of the individual ones are found on the real
axis, to treat them separately, it is convenient to introduce the total
resolvent,
\begin{equation}
 \omega(x) = \sum_{r=0}^{k-1} \omega^{(r)}
  \left(x - \frac{2\pi i}{k}r \right).
\end{equation}
Fig.~\ref{cuts} shows the singularities of the total resolvent.
Their boundary conditions are given by
\begin{equation}
 \omega^{(r)}(x) \longrightarrow \pm T, \qquad
 \omega(x) \longrightarrow \pm kT \qquad
 \mbox{as} \qquad x \longrightarrow \pm \infty.
\end{equation}
In the large $N$ limit these resolvents can be also represented with
the distribution functions as
\begin{eqnarray}
 \omega^{(r)}(x) & = & T \int dy \rho^{(r)}(y) 
  \coth \left( \frac{x-y}{2} \right),
 \\
 \omega(x) & = & T \sum_{r=0}^{k-1} \int dy \rho^{(r)}(y) 
  \coth \left( \frac{x-y}{2} - \frac{\pi i}{k}r \right).
\end{eqnarray}
Corresponding to this total resolvent, it is convenient to define the
total density of eigenvalues,
\begin{equation}
 \rho(x) = \sum_{r=0}^{k-1} \rho^{(r)}
  \left(x - \frac{2\pi i}{k}r \right).
\end{equation}
This has supports not only on the real axis but also parallel lines as
the total resolvent.
Thus we can rewrite the resolvent, the prepotential and the saddle
point equation in a simple form as
\begin{equation}
 \omega(x) = T \int dy \rho(y) \coth \left( \frac{x-y}{2} \right),
\end{equation}
\begin{equation}
 \mathcal{F} = T \sum_{r=0}^{k-1} \int dx \rho^{(r)}(x) V(x)
  - T^2 \PV \int dx dy \rho(x) \rho(y)
  \log \sinh \left( \frac{x-y}{2} \right),
\end{equation}
\begin{eqnarray}
 \frac{1}{T} V'(x) & = & 
  \PV \int dy \rho(y) 
  \coth \left(\frac{x-y}{2}+\frac{\pi i}{k}r\right).
  \label{saddle02}
\end{eqnarray}
From multi-cut discontinuities of the resolvents we obtain the
density of eigenvalues
\begin{eqnarray}
 \rho^{(r)}(x) & = & - \frac{1}{4\pi i T}
  \left(
   \omega^{(r)}(x+i\epsilon) - \omega^{(r)}(x-i\epsilon)
  \right)
\end{eqnarray}
which has supports on the intervals $\mathcal{C}^{(l,r)}$.
Similarly we have
\begin{equation}
 \frac{1}{2T}
  \left(
   \omega^{(r)}(x+i\epsilon) + \omega^{(r)}(x-i\epsilon)
  \right)
  = \PV \int dy \rho^{(r)}(y)
  \coth \left(\frac{x-y}{2}\right).
\end{equation}
Therefore the saddle point equations (\ref{saddle02}) can be written as
\begin{equation}
 V'(x) = \frac{1}{2}
  \left[
    \omega\left(x+\frac{2\pi i}{k}r+i\epsilon\right)
   +\omega\left(x+\frac{2\pi i}{k}r-i\epsilon\right)
  \right] .
  \label{saddle03}
\end{equation}
To discuss geometric aspects of the theory, it is convenient to introduce
a new function
\begin{equation}
 y^{(r)}(x) = V'(x) - \omega \left( x + \frac{2\pi i}{k}r \right) 
  = {V^{(r)}_{\rm eff}}'(x).
\end{equation}
%The saddle point equation can be written in a simple way as
%\begin{equation}
% y^{(r)}(x) = 0.
%\end{equation}
When we divide the resolvent into regular and singular parts $\omega(x)
= \omega_{\rm reg}(x) + \omega_{\rm sing}(x)$, due to the saddle point
equation, this function is given by
\begin{equation}
 y^{(r)}(x) = - \omega_{\rm sing} \left( x + \frac{2\pi i}{k}r \right).
\end{equation}
We also introduce another function as well as the total resolvents,
defined as
\begin{equation}
 y(x) = k V'(x) - \omega(x).
\end{equation}
This function has cuts $\tilde\mathcal{C}^{(l,r)}=\{x+2r\pi i/k | x \in
\mathcal{C}^{(l,r)}, l=1, \cdots, n, r = 0, \cdots, k-1\}$ as shown in
Fig.~\ref{cuts}, corresponding to all the cuts of the original
individual resolvents $\omega^{(r)}(x)$.
Since the resolvent is periodic in the imaginary direction, $x
\sim x + 2\pi i$, there seems to be the discrete $\Z_k$ shift
symmetry $x \to x + 2\pi i/k$.
This is quite similar to the discrete $\Z_n$ symmetry of $\SU(n)$
five dimensional theory, proposed in \cite{Nekrasov:1996cz}.
This similarity suggests the level-rank duality of this model.

Thus the partial 't Hooft coupling is simply given by its contour integral,
\begin{equation}
  T^{(l,r)} 
   = \frac{1}{4\pi i} \oint_{\mathcal{C}^{(l,r)}} 
   dx \ \omega_{\rm sing}\left(x+\frac{2\pi i}{k}r\right)
   = - \frac{1}{4\pi i} \oint_{\tilde\mathcal{C}^{(l,r)}} 
   dx \ y(x).
\end{equation}
In addition, from (\ref{legendre_part}) we have another contour integral
\begin{equation}
 \frac{\partial \mathcal{F}}{\partial T^{(l,r)}}
 = - \Gamma^{(l,r)}
 = - \frac{1}{2} \oint_{\mathcal{D}^{(l,r)}} 
  dx \ y(x),
\end{equation}
where the contour $\mathcal{D}^{(l,r)}$ is the relative cycle
represented by the path starting at $\infty_+$, and going back to $\infty_-$
on another sheet after reaching $\tilde\mathcal{C}^{(l,r)}$ as shown in
Fig.~\ref{cuts}.
Indeed this relation is analogous to the relation for Seiberg-Witten curve
\begin{equation}
 a_l = \oint_{A_l} dS, \qquad
 \frac{\partial \mathcal{F}}{\partial a_l} = a^D_l = \oint_{B_l} dS .
\end{equation}
For our matrix model it is shown in the following discussion that
the one-form obtained from the spectral curve of the matrix model coincides
with Seiberg-Witten differential.

\subsection{Relation to Seiberg-Witten theory}

We now discuss the relation between Seiberg-Witten curve and the matrix model.
In the first place, the matrix model captures the asymptotic behavior
of the combinatorial representation of the partition function.
The energy functional, which is derived from the asymptotics of the
partition function \cite{Nekrasov:2003rj}, in terms of the profile function 
\begin{equation}
 \mathcal{E}_\Lambda(f) = \frac{1}{4} \PV \int_{y<x} dx dy
  f''(x) f''(y) (x-y)^2
  \left(
   \log \left(\frac{x-y}{\Lambda}\right) - \frac{3}{2}
  \right)
  \label{functional01}
\end{equation}
can be rewritten as
\begin{equation}
 {\rm E}_\Lambda(\varrho) = - \PV \int_{x\not=y} dx dy
  \frac{\varrho(x)\varrho(y)}{(x-y)^2}
  -2 \int dx \varrho(x) \log \prod_{l=1}^N
  \left(\frac{x-a_l}{\Lambda}\right),
  \label{functional02}
\end{equation}
up to the perturbative contribution
\begin{equation}
 \frac{1}{2} \sum_{l,m} (a_l-a_m)^2 
  \log \left(\frac{a_l-a_m}{\Lambda}\right),
\end{equation}
by identifying
\begin{equation}
 f(x) - \sum_{l=1}^n |x-a_l| = \varrho(x).
\end{equation}
Then integrating (\ref{functional02}) by parts, we have
\begin{equation}
 {\rm E}_\Lambda (\varrho) = 
  - \PV \int_{x\not=y} dx dy \varrho'(x) \varrho'(y) \log (x-y)
  + 2 \int dx \varrho'(x) \sum_{l=1}^n
  \left[(x-a_l)\log\left(\frac{x-a_l}{\Lambda}\right)-(x-a_l)\right].
\end{equation}
This is just the matrix model with the potential (\ref{pot_vec4d}) with $k=1$ if
we identify $\varrho'(x)=\rho(x)$.
Therefore analysis of this matrix model is equivalent to that of
\cite{Nekrasov:2003rj}.
But in this section we reconsider the result of the gauge theory from
the viewpoint of the matrix model (similar approach is found in
\cite{Marshakov:2011vw}).

\subsubsection{Four dimensional theory}

Let us first consider the most simple case, the four dimensional theory with
$k=1$, as a preliminary for other generalized theories.
This is the case investigated in \cite{Nekrasov:2003rj} in detail.
We now concentrate on the relation between Seiberg-Witten curve and the
spectral curve of the matrix model.
One can see notations for the four dimensional matrix model in appendix
\ref{sec:matrix4d}.

To clarify the connection between the matrix model description and the
gauge theory consequences explicitly, we have to consider the
contribution of the dynamical scale, which is actually interpreted as the
instanton counting parameter.
In this case the counting parameter should be replaced with
$\Lambda^{(2n-N_f)|\vec{\lambda}|/k}$ because the instanton number
becomes fractional \cite{Fucito:2004ry}.
In this section we introduce this contribution to the potential functions
for both the vector multiplet and the (anti)fundamental matter.

Due to the saddle point equation (\ref{EOM4d}), we have a regular
function on the complex plane, except at the infinity,
\begin{equation}
 P_n(z) = \Lambda^n \left(e^{y/2} + e^{-y/2} \right)
  \equiv \Lambda^n \left(w + \frac{1}{w}\right).
  \label{elliptic01}
\end{equation}
This turns out to be a monic polynomial $P_n(x) = z^n + \cdots$,
because it is an analytic function with the following asymptotic behavior, 
\begin{equation}
 \Lambda^n e^{y/2} 
= \Lambda^n e^{-\omega(z)} \prod_{l=1}^n \left(\frac{z-a_l}{\Lambda}\right)
  \longrightarrow z^n , \qquad
  z \longrightarrow \infty .  
\end{equation}
Here $w$ should be the smaller root with the boundary condition as
\begin{equation}
 w \longrightarrow \frac{\Lambda^n}{z^n},
  \qquad
  z \longrightarrow \infty, 
\end{equation}
thus we now identify
\begin{equation}
 w = e^{-y/2} .
\end{equation}
Therefore from the hyperelliptic curve (\ref{elliptic01}) we can relate
Seiberg-Witten curve to the spectral curve of the matrix model,
\begin{equation}
 dS = \frac{1}{2\pi i} z \frac{dw}{w} 
  = - \frac{1}{2\pi i} \log w \ dz
  = \frac{1}{4 \pi i} y (z) dz.
\end{equation}
Note that it is shown in \cite{Marshakov:2006ii,Marshakov:2011vw} we
have to take the vanishing fraction limit to obtain the Coulomb moduli from the
matrix model contour integral.
This is the essential difference between the profile function method and
the matrix model description.

We can deal with the theory with (anti)fundamental matters in a similar way.
In this case the spectral curve of the matrix model reads
\begin{equation}
 e^{y/2} = \frac{1}{\Lambda^{n-N_f/2}} e^{-\omega} 
  \prod_{l=1}^n (x-a_l) \prod_{f=1}^{N_f} (x+m_f)^{-1/2}
  \stackrel{x\to\infty}{\longrightarrow} \frac{1}{\Lambda^{n-N_f/2}} x^{n-N_f/2}.
\end{equation}
Thus the regular part is modified as
\begin{equation}
 \frac{P_n(z)}{\sqrt{Q(z)}} = \Lambda^{n-N_f/2} \left(e^{y/2}+e^{-y/2}\right)
  \equiv \Lambda^{n-N_f/2} \left( w + \frac{1}{w} \right) ,
\end{equation}
where we now take into account the contribution from the matter fields,
\begin{equation}
 Q(z) = \prod_{f=1}^{N_f} (x + m_f).
\end{equation}
As well as the pure gauge theory, we identify $w = e^{-y/2}$ and $P_n(z)$
is a monic polynomial because of its analyticity and asymptotics.

For $k>1$ although we have to consider the multi-matrix model, we can
perform almost the same approach.
In this case we now apply the same resolvent $\omega$ and the
corresponding function $y$ to all the matrices.
Due to the slightly modified potential term, we have a regular
function
\begin{equation}
 P_n(z) 
  = \Lambda^n \left(e^{ky/2} + e^{-ky/2} \right)
  \equiv \Lambda^n \left(w + \frac{1}{w}\right).
\end{equation}
Again $P_n(z)$ is a monic polynomial $P_n(z) = z^n + \cdots $, due to
the asymptotic behavior,
\begin{equation}
 \Lambda^n e^{ky/2} = \Lambda^n e^{-k\omega(z)}
  \prod_{l=1}^n \left(\frac{z-a_l}{\Lambda}\right)
  \stackrel{z \to \infty}{\longrightarrow} z^n.
\end{equation}
Identifying the smaller root as
\begin{equation}
 w = e^{-ky/2} \stackrel{z\to\infty}{\longrightarrow} 
  \frac{\Lambda^{n}}{z^n},
\end{equation}
we have a slightly modified relation for the spectral curve 
\begin{equation}
 \frac{1}{4\pi i}y(z) dz = \frac{1}{2k\pi i} z \frac{dw}{w}.
\end{equation}
We can apply the same spectral curve to each matrix because
they are independent of the index $r$ of the matrix.
Therefore Seiberg-Witten curve for each matrix is given by multiplying
the original one by $1/k$,
\begin{equation}
 dS \longrightarrow \frac{1}{k} dS.
  \label{SWcurve4d}
\end{equation}
It means Seiberg-Witten curve for the ALE space is decomposed to $k$
curves, and when we consider the total contribution from $k$ matrices, we can
reproduce the original relation for the prepotential and the Coulomb moduli.
We can easily see the relation (\ref{SWcurve4d}) is also obtained from the
theory with the matter fields.

This simple relation between $k=1$ and $k>1$ theories is also suggested
by the result in \cite{Nishioka:2011jk}.
It is shown that the central charge of the two dimensional conformal
field theory, corresponding to the theory we discuss here, is given by
\begin{equation}
 c = kn + \frac{n^3-n}{k} \frac{(\epsilon_1+\epsilon_2)^2}{\epsilon_1
  \epsilon_2}.
\end{equation}
In the case of $\beta=1$, namely $\epsilon_2 = -\epsilon_1$, since the second
part is vanishing, it is natural to expect there are $k$ sectors without
interacion between them.
This is consistent with our result discussed above.

\subsubsection{Five dimensional theory}

We generalize the result for $k=1$, which is derived in
\cite{Nekrasov:1996cz} (see also \cite{Nekrasov:2003rj}), to the cases $k>1$.
We now apply a similar method discussed above to the five dimensional case.
As well as the four dimensional theory, we can have a regular function
\begin{equation}
 X^{-kn/2} P_{kn}(X) = \Lambda^{n}
  \left( e^{y/2} + e^{-y/2} \right)
  \equiv \Lambda^n \left( w+\frac{1}{w} \right),
  \qquad X = e^z.
  \label{SWcurve5d}
\end{equation}
where we modify the definition $y \to  y + kT$.
In this case it yields a $kn^{\rm th}$ monic polynomial $P_{kn}(X) =
X^{kn} + \cdots$, because of its asymptotics
\begin{equation}
 \Lambda^n e^{y/2} = \Lambda^n e^{(kT-\omega(z))/2} \prod_{l=1}^n
  \left(\frac{2}{\Lambda} \sinh \frac{k}{2}(z-a_l)\right)
  \stackrel{X\to\infty}{\longrightarrow} X^{kn/2}.
\end{equation}
By setting the boundary condition
\begin{equation}
 w = e^{-y/2} \stackrel{X\to\infty}{\longrightarrow} 
  \frac{\Lambda^n}{X^{kn/2}},
\end{equation}
Seiberg-Witten differential for this theory defined on the
curve (\ref{SWcurve5d}) reads
\begin{equation}
 dS
  = \frac{1}{4\pi i} y(z) d z
  = \frac{1}{2\pi i} \log (X) \frac{dw}{w}.
\end{equation}
We can see the result of \cite{Nekrasov:1996cz} is obtained by setting $k=1$.
In this case although the curve is modified as (\ref{SWcurve5d}), the
form of the differential itself is not changed.
In the four dimensional limit $R \to 0$, $k$ sets of $n$ cuts are
decoupled because distance between them is $2 \pi i/k R$ as shown in Fig.~\ref{cuts}.
A generalization to the theory with matter fields is straightforward.

\section{Further research}\label{sec:related}

The recent remarkable progress on the four dimensional $\N=2$ theory
gives the interesting relation to the two dimensional conformal field theory.
Thus, according to the results discussed above, it is natural to search a two
dimensional theoretical counterpart of the orbifold theory.
In this section we discuss some proposals for such an interesting 2d/4d
connection.

\subsection{String theory perspective}

We now discuss a string theory realization of the orbifold theory, and
try to find its two dimensional description.
It is well known that string theory gives us a lot of interesting field
theoretical consequences, e.g. the holographic approach to a strongly
correlated system, the method to construct instantons and monopoles and so on.
Actually the 2d/4d relation \cite{Alday:2009aq} can be naturally
understood in terms of M-theory.

Let us consider the string theory realization of $\N=4$ SYM theory on the
Taub-NUT space \cite{Dijkgraaf:2007sw}, which is expected to be related
to our case.
The Taub-NUT manifold $TN_k$ is given by a $S^1$ compactification of the
$A_{k-1}$ singularity.
It approaches to the cylinder $\R^3 \times S^1$ at infinity, and 
the ALE space is obtained by taking the limit of $R\to\infty$ where $R$
stands for the radius of $S^1$.
We remark although the partition function of the gauge theory on the Taub-NUT
manifold is not identical to that of the ALE space, we expect this
difference does not seriously affect the corresponding two dimensional
theory because the $\SU(n)$ theory on both of these manifolds is
specified by a flat connection on the boundary $S^3/\Z_k$ as discussed later.

We start with type IIA string theory on $TN_k \times S^1 \times \R^5$
with wrapping $n$ D4-branes on $TN_k \times S^1$.
Lifting this to M-theory, we then obtain the compactification of $TN_k
\times T^2 \times \R^5$ and wrapping $n$ M5-branes on $TN_k \times T^2$.
By replacing the two-manifold $T^2$ with various Riemann surfaces $\Sigma$, we
obtain the corresponding $\N=2$ theories \cite{Gaiotto:2009we}.
To study this configuration we go back to type IIA theory with another
compactification $\R^3 \times T^2 \times \R^5$.
In this case there are $n$ D4-branes and $k$ D6-branes wrapping $\R^3
\times T^2$ and $T^2 \times \R^5$ respectively because the circle
fibration of the Taub-NUT space has singular points.
One can see these D4- and D6-branes are intersecting on $T^2$, and thus
the chiral fermion is arising from these intersecting configuration.
This chiral fermion plays an important role in considering the
level-rank duality of the system.

To discuss the two dimensional theory on $T^2$ we then deal with the
boundary of the four dimensional manifold.
By considering the radial quantization near the boundary and a
wavefunction for the time evolution along $S^3/\Z_k \times \R$, we have
a Hilbert space with a state $|\rho\rangle$ for each $n$-dimensional
representation
\begin{equation}
 \rho : \ \Z_k \subset \SU(2) \longrightarrow \SU(n).
  \label{embedding01}
\end{equation}
Integrating out the flux coming from $k$ D6-branes, we
obtain the Chern-Simons term for 
\begin{equation}
 I_{\rm CS} = 2 \pi k \int_{T^2\times\R} CS(A).
\end{equation}
This means the boundary condition for the D4-brane requires specifying a
state of the $\SU(n)$ Chern-Simons theory at level $k$ living on $T^2$,
and thus we have a state for each integrable representation of the
$\widehat\SU(n)_k$ WZW model.
Note that the same procedure can be performed for the $\SU(k)$ theory on
$k$ D6-branes, and the diagonal $U(1)$ part is decoupled.
Therefore we obtain the embedding $\widehat\U(1)_{nk} \times
\widehat\SU(n)_k \times \widehat\SU(k)_{n} \subset \widehat\U(nk)_1$,
thus it is easy to see the level-rank duality.

We also remark it is conjectured in \cite{Nekrasov:2002qd} that the
gauge theory partition function is related to the 
$\tau$-function by utilizing the chiral fermion representation.
Consequently the generating function for instantons on the ALE space of
the ADE type would be related to the ADE WZW theories on the
Seiberg-Witten curve.
It is consistent with the intersecting brane configuration because we
obtain the $\N=2$ theories when we consider the intersecting branes on
general curves $\Sigma$, which turn out to be Seiberg-Witten/Gaiotto
curves \cite{Gaiotto:2009we}.

We then discuss the connection between the two and four dimensional
theory more concretely.
The original proposals of \cite{Alday:2009aq,Wyllard:2009hg} elucidate
the explicit relation between the $\SU(n)$ partition function and
the two dimensional conformal field theory described by $W_{n}$
algebra.
This relation is extended to the theory with the generalized $W$ algebra
\cite{Alday:2010vg,Kozcaz:2010yp} \cite{Tachikawa:2011dz}
\cite{Kanno:2011fw}, which is obtained from the quantum Drinfeld-Sokolov
reduction applied to an affine Lie algebra $\widehat\SU(n)$
\cite{deBoer:1993iz}.
It is characterized by a choice of an embedding $\rho : \SU(2) \to
\SU(n)$, and this embedding $\rho$ can be labeled by a partition of $n$, or
equivalently a Young diagram $Y$.
For example, it reproduces well-known algebras,
$W(\widehat\SU(n),[1,\cdots,1])=\widehat\SU(n)$,
$W(\widehat\SU(n),[N])=W_n$, etc.
Such an embedding has been discussed in the orbifold case
(\ref{embedding01}).
Indeed we consider the embedding of the finite subgroup $\Z_k$ of
$\SU(2)$ for the orbifold theory, and the decomposition into the
irreducible representations of $\Z_k$ corresponds to that of the gauge
group $\SU(n) \to \SU(n_0) \times \cdots \times \SU(n_{k-1})$ with
$n_0+\cdots+n_{k-1}=n$.
This shows that this embedding is also labeled by a partition of $n$.
Therefore it is expected that the two dimensional conformal field
theory, corresponding to the $\SU(n)$ gauge theory on the ALE space
$\C^2/\Z_k$, is related to the generalized $W$ algebra
$W(\widehat\SU(n),[n_0,\cdots,n_{k-1}])$.
This connection is still speculation, and should be investigated in detail.

\subsection{$q$-deformed CFT}

It is shown in \cite{Awata:2009ur} that $q$-deformed CFT, which
is described by $q$-Virasoro \cite{Shiraishi:1995rp} and
$q$-$W$ algebra \cite{Awata:1995zk}, corresponds to the
$q$-deformed five dimensional partition function of the gauge theory,
and its matrix model description is also proposed in \cite{Awata:2010yy}
by clarifying a connection with $q$-Virasoro algebra.
This matrix model, which we now call $q$-Virasoro matrix model, is apparently
different from the trigonometric one
\cite{Klemm:2008yu,Sulkowski:2009br,Sulkowski:2009ne}, but it is worth
considering the root of unity limit of such a $q$-Virasoro related
models.
The root of unity limit of $q$-Virasoro algebra is investigated in
\cite{Bouwknegt1997qru} while that of $q$-$W$ algebra
is not yet well known.

In this subsection let us focus on the matrix measure part of the
$q$-Virasoro matrix model \cite{Awata:2010yy}.
It is just given by the two-parameter deformed Vandermonde determinant,
which is closely related to the Macdonald polynomial \cite{Mac_book},
\begin{equation}
 \Delta_{q,t}^2(x) = \prod_{i\not=j} 
  \frac{(x_i/x_j;q)_\infty}{(tx_i/x_j;q)_\infty}.
\end{equation}
We can obtain the usual Vandermonde determinant by taking the limit $q
\to 1$ with $t=q^\beta$,
\begin{equation}
 \Delta_{q,t}^2(x) \longrightarrow \prod_{i\not=j}
  \left(
   1 - \frac{x_i}{x_j}
  \right)^\beta
  \simeq \prod_{i<j} (x_i - x_j)^{2\beta}.
\end{equation}
This corresponds to the limit to get the Jack polynomial from the
Macdonald polynomial \cite{Mac_book}.
On the other hand, performing the Uglov condition, $q \to \omega q$, $t
= \omega q^\beta$ with $\beta = k \gamma + 1 \equiv 1$ $(\mbox{mod}\
k)$, then taking $q \to 1$, it becomes
\begin{eqnarray}
 \Delta_{q,t}^2(x) \longrightarrow \prod_{i\not=j}
  \left(
   1 - \frac{x_i}{x_j}
  \right)
  \left(
   1 - \frac{x_i^k}{x_j^k}
  \right)^\gamma
  &\simeq &
  \prod_{i<j} (x_i - x_j)^2 (x_i^k - x_j^k)^{2\gamma}
  \nonumber \\
  & = & \prod_{i<j} (x_i - x_j)^{2+2\gamma} 
  ( x_i^{k-1} + x_i x_j^{k-2} + \cdots + x_j^{k-1})^{2\gamma}.
  \nonumber \\
 \label{vander02}
\end{eqnarray}
The matrix measure (\ref{vander02}) is expected to be related to the
 corresponding one (\ref{vander4d}).
Indeed they are equivalent for $k=1$.
Although this is written in terms of only one matrix while
the model discussed in the previous section is not, we can see the
 Vandermonde part in (\ref{vander02}) is also found in (\ref{vander4d}).
We just expect the interaction part in (\ref{vander4d}) is encoded into
 the non-singular part of (\ref{vander02}).

Such a difference seems to be related to that found in the usual matrix
model for the four dimensional theory, the Penner-type
\cite{Dijkgraaf:2009pc,Eguchi:2009gf,Schiappa:2009cc,Eguchi:2010rf},
interpreted as the integral representation of the Liouville correlator and
the other one just given by the asymptotics of the combinatorial
expression of the partition function
\cite{Klemm:2008yu,Sulkowski:2009br,Sulkowski:2009ne}.
Indeed they are different because the conformal symmetry is manifest
for the former one, but the latter one.
However it is not enough yet to understand the difference between the
Macdonald-type and trigonometric-type deformed Vandermonde determinant.
This difference is expected to be understood, for example, in terms of
string theory.

We now remark the polynomial, called the {\it Uglov polynomial} in
\cite{KuramotoKato200908}, is given by taking this limit from the
Macdonald polynomial in order to describe the spin Calogero-Sutherland
model \cite{Uglov:1997ia}.

\section{Summary and discussion}\label{sec:summary}

In this paper we have performed some extensions of the partition
function and its matrix model description for the four dimensional
$\N=2$ gauge theory to the orbifold theory.
We have shown that the orbifold projection, extracting the
$\Gamma$-invariant sector of the Young diagram, is automatically
performed by taking the root of unity limit $q \to \exp \left(2\pi i /
k\right)$ of the $q$-deformed partition function while we have to
take $q \to 1$ to get the four dimensional theory on $\R^4$.
For such an orbifold partition function of $\SU(n)$ theory with
$\Gamma=\Z_k$, it is convenient to divide $n$-tuple to
$kn$-tuple partitions.
As a result, we have obtained the multi-matrix model by considering the
asymptotic behavior of the combinatorial representation.
This matrix model at the large $N$ limit has been analysed in detail,
which is equivalent to studying the limit shape of the Young diagram,
and then we have seen Seiberg-Witten curve is obtained as the spectral
curve of the matrix model.
We have also discussed the corresponding two dimensional theory of the
four dimensional orbifold theory.
Focusing on the embedding $\rho: \Z_k \subset \SU(2) \to \SU(n)$, which
characterizes the decomposition of $\SU(n)$ gauge group to the irreducible
representations of $\Z_k$, we have suggested the generalized $W$ algebra
appears in the two dimensional theory.

We now comment on some possibilities of extension beyond this study.
We hope our study becomes a step for understanding of M-theory itself.
Actually the emergence of $\widehat\SU(n)_k$ symmetry in the two dimensional
theory as a counterpart of the orbifold theory $\C^2/\Z_k$ is analogous
to the ABJM theory because
the level of Chern-Simons theory in the ABJM theory is directly related
to the background manifold $\C^4/\Z_k$ on which M2-branes are located.
Indeed, in both cases, the level $k$ is interpreted as the degree of
singularity of the complementary manifold of the corresponding
world-volume theory.
It is interesting to study a relation between M2- and M5-branes from
this point of view.
We are also interested in the level-rank duality of our model.
Such a duality can be found in the two dimensional $\widehat\SU(n)_k$, or
$\widehat\SU(k)_n$ WZW theory, which is
closely related to the four dimensional orbifold theory.
It is expected that this duality plays an important role in
understanding some aspects of M-theory.

It is natural to consider some applications to related topics.
One of them is the three dimensional duality, which is a recent hot
topic on this subject
\cite{Terashima:2011qi,Dolan:2011rp,Galakhov:2011hy,Gadde:2011ia,Gadde:2011ik,Imamura:2011uw,Nishioka:2011dq}.
The $q$-parameter plays a similar role in such a theory, so that it is
interesting to study the singular limit of the $q$-parameter, i.e. the
root of unity limit.
Second is a relation to the quantized integrable models.
It is shown in \cite{Nekrasov:2009rc} that when we consider generic
$\Omega$-parameters, $\epsilon_1 + \epsilon_2 \not= 0$, correction
to the prepotential can be interpreted as a quantization effect of the
corresponding integrable model.
Searching an integrable model, which corresponds to the orbifold theory,
would be also interesting.

%%%%%%%%%  Acknowledgments  %%%%%%%%%

\subsection*{Acknowledgments}
The author would like to thank S.~Hikami for reading the manuscript and
useful comments.
The author also would like to thank T.~Azeyanagi, K.~Hashimoto, Y.~Kato,
K.~Maruyoshi, H.~Shimada, T.~Tai and M.~Taki for valuable discussions
and comments.
This work is supported by Grant-in-Aid for JSPS Fellows.% (No.~23-593).

\appendix
\section{Four dimensional matrix model}\label{sec:matrix4d}

In this appendix we summarize the results of the four dimensional limit
of the matrix model in order to fix our notations.
Let us mainly consider the $k=1$ theory, and shortly comment on simple
generalizations to the cases of $k>1$.

The partition function for this case is given by
\begin{equation}
 Z = \int \prod_{i=1}^N \frac{dx_i}{2\pi}
  \prod_{i<j}^N (x_i-x_j)^2
  e^{-\frac{1}{g_s}\sum_{i=1}^N V(x_i)},
\end{equation}
and the saddle point equation is obtained from its differentiation with
respect to each eigenvalue,
\begin{equation}
 V'(x_i) = 2 g_s \sum_{j(\not=i)}^N \frac{1}{x_i-x_j}.
\end{equation}
This is also given by the extremal condition of the effective potential
defined as
\begin{equation}
 V_{\rm eff}(x_i) = V(x_i) - 2 g_s \sum_{j(\not=i)}^N \log(x_i-x_j).
\end{equation}

We then introduce the resolvent for this model.
By taking the large $N$ limit, it can be given by the integral representation,
\begin{equation}
 \omega(x) = T \int dy \frac{\rho(y)}{x-y}.
\end{equation}
Its asymptotic behavior yields
\begin{equation}
 \omega(x) \longrightarrow \frac{1}{x}, \qquad
  x \longrightarrow \infty.
\end{equation}
Here the density of states is obtained from the discontinuities of the
resolvent,
\begin{equation}
 \rho(x) = - \frac{1}{2\pi i T} 
  \left(\omega(x+i\epsilon) - \omega(x-i\epsilon)\right).
\end{equation}
Thus the saddle point equation can be also written in the following
form, which is convenient to discuss its analytic property,
\begin{equation}
 V'(x) = \omega(x+i\epsilon) + \omega(x-i\epsilon).
  \label{EOM4d}
\end{equation}

On the other hand, we have another convenient form to treat the saddle
point equation, which is called the loop equation, given by
\begin{equation}
 y^2(x) - {V'(x)}^2 + R(x) = 0
\end{equation}
where we denote
\begin{eqnarray}
  y(x) & = & V'(x) - 2\omega(x) = -2 \omega_{\rm sing}(x), 
   \nonumber \\
 R(x) & = & \frac{4T}{N} \sum_{i=1}^N 
  \frac{V'(x)-V'(x_i)}{x-x_i}.
\end{eqnarray}
It is obtained from the saddle point equation by multiplying $1/(x-x_i)$
and taking their summation and the large $N$ limit.
This is interpreted as the hyperelliptic curve which is given by
resolving the singular form,
\begin{equation}
 y^2(x) - {V'(x)}^2 = 0 .
\end{equation}
The genus of the Riemann surface is directly related to the number of
cuts of the corresponding resolvent.
The filling fraction, or the partial 't Hooft coupling, is simply given
by the contour integral on the hyperelliptic curve
\begin{equation}
 T_l = \frac{1}{2\pi i} \oint_{\mathcal{C}_l} dx \ \omega_{\rm sing}(x)
  = - \frac{1}{4\pi i} \oint_{\mathcal{C}_l} dx \ y(x).
\end{equation}

For $k>1$ we can perform the same thing as well as the case of $k=1$
since there is no interaction between $k$ matrices in the four
dimensional limit with $\beta=1$.
The $k$-matrix partition function is given by
\begin{equation}
 Z = \prod_{r=0}^{k-1} Z^{(r)},
\end{equation}
\begin{equation}
 Z^{(r)} = \int \prod_{i=1}^N \frac{dx^{(r)}_i}{2\pi}
  \prod_{i<j}^N (x^{(r)}_i-x^{(r)}_j)^2
  e^{-\frac{1}{g_s}\sum_{i=1}^N V(x^{(r)}_i)}.
\end{equation}
We can provide all the matrices with the same density of states, resolvents and
hyperelliptic curves etc, e.g. $y^{(r)}(z)$, and the total contribution
to the partition function is given by the simple summation of them.
On the other hand, when we consider the case $\beta\not=1$, we have to
take into account interaction between matrices.
It is a quite interesting situation, and will be investigated in a
future work.

%%%%%%%%%  Rerefence  %%%%%%%%%

\providecommand{\href}[2]{#2}\begingroup\raggedright\endgroup

\end{document}